# Cauchy-Maxwell equations: A unified field theory for coupled electromagnetism and elasticity

Pranesh Roy[1,§], Sanjeev Kumar[1,2] and Debasish Roy[1,2,*]

[1]*Computational Mechanics Lab, Department of Civil Engineering*

*Indian Institute of Science, Bangalore 560012, India*

[2]*Centre of Excellence in Advanced Mechanics of Materials*

*Indian Institute of Science, Bangalore 560012, India*

*([*]Corresponding author; email: royd@iisc.ac.in)*

## Abstract

A conformal gauge theory is used to describe and unify myriad electromechanical and magnetomechanical coupling effects observed in solid continua. Using a space-time pseudo-Riemannian metric in a finite-deformation setup and exploiting the Lagrangian's local conformal symmetry, we derive Cauchy-Maxwell (CM) equations that seamlessly combine, for the first time, Cauchy's elasto-dynamic equations with Maxwell's equations for electromagnetism. Maxwell's equations for vacuum are recoverable from our model, which in itself also constitutes a new derivation of these equations. With deformation gradient and material velocity coupled in the Lagrange density, various pseudo-forces appear in the Euler-Lagrange equations. These forces, not identifiable through classical continuum mechanics, may have significance under specific geometric or loading conditions. As a limited illustration on how the CM equations work, we carry out semi-analytical studies, viz. on an infinite body subject to isochoric deformation and a finite membrane under both tensile and transverse loading, considering piezoelectricity and piezomagnetism. Our results show that under specific loading frequencies and tension, electric and magnetic potentials may increase rapidly in some regions of the membrane. This may have significance in future studies on efficient energy harvesting.

**Keywords**: gauge theory, conformal symmetry, electro-magneto-mechanical response, piezoelectricity, piezomagnetism

---

[§] *Current Address: Department of Aerospace and Mechanical Engineering, University of Arizona, Tucson, AZ 85721, USA*

## 1 Introduction

With their far-reaching industrial relevance, electromechanical and magnetomechanical coupling phenomena have been of interest since the invention of piezoelectricity by Jacques and Pierre Curie in 1880. A few polymers (e.g. polyvinylidene fluoride (PVDF)), ceramics (e.g. lead zirconate titanate (PZT)) and many naturally occurring materials exhibit piezoelectricity. Although piezoelectricity is the most technologically exploited and extensively investigated phenomenon of electromechanical coupling, several other forms of electro-magneto-mechanical coupling exist and are used in various applications. Notable among them is flexoelectricity - a coupling of strain gradient with polarization that exists in all dielectric materials and of relevance in the nanoscale (see Maranganti *et al.*, 2006 and Majdoub *et al.*, 2008). A few materials show high flexoelectric coefficients, e.g. polycrystalline centrosymmetric perovskites such as $\left(\mathrm{Ba,Sr}\right)\mathrm{TiO_3}$, ferroelectric, incipient ferroelectric and relaxor ferroelectric perovskites. paraelectric $\mathrm{SrTiO_3}$, a few liquid crystals, mechano-sensitive bio-membranes and polymers. Electrostriction is an electromechanical coupling phenomenon observed in all dielectric materials to an extent, wherein strain is coupled with the quadratic of polarization (Newnham *et al.*, 1997). A few materials show relatively high electrostriction, e.g. Lead Magnesium Niobate, Lead Lanthanum Zirconate Titanate, Barium Titanate, electrostrictive polyurethane etc. Prominent among various magnetomechanical coupling phenomena is flexomagnetism - a magnetoelastic coupling between strain gradient and magnetization, which is expected to be pronounced in the nanoscale in $\mathrm{Mn}$-based antiperovskites, e.g. $\mathrm{Mn_3GaN}$ (Lukashev and Sabirianov, 2010). Piezomagnetism constitutes a linear coupling of strain and magnetization occurring in some antiferromagnetic materials e.g. $\mathrm{MnF_2}$, $\mathrm{FeF_2}$ and $\mathrm{CoF_2}$ (Borovik-Romanov, 1960 and Moriya, 1959). Magnetostriction implies a magnetomechanical coupling, i.e. strain is coupled with the quadratic of magnetization and is observed in ferromagnetic materials e.g. Terfenol-D, Alfenol, Galfenol, Metglas, Cobalt ferrite etc. (Sandlund *et al.*, 1994). In order to predict the response of materials exhibiting electromechanical and magnetomechanical coupling, many theoretical models have been suggested (see Tiersten, 2013, Yang and Suo, 1994, Richards and Odegard (2010), Guo and Guo, 2003, Furthmüller *et al.*, 1987, DeSimone and James, 1997, James and Kinderlehrer, 1994, Lukashev and Sabirianov, 2010, Newnham *et al*., 1997, Roy and Roy 2019). These are based on an empirically introduced energy density functional and consider small deformation and static loading conditions only. Predicting the response of these materials under large deformation and dynamic conditions remains a challenge. In order to make progress, one needs to find a consistent way to couple Cauchy's elasticity equations with Maxwell's equations of electromagnetism. In this work, we report an attempt in that direction.

To make progress, we wish to extend the quasi-static formulation attempted in a previous article by Roy *et al*. (2019) to the dynamic case. In the last cited work, we had developed a conformal gauge theory of solids to model various electromechanical and magnetomechanical coupling phenomena under static loading conditions. Considering that a scaling of the right Cauchy-Green tensor (the pulled back metric) by a constant factor leaves the isochoric part of the Lagrange density invariant with only the volumetric part getting affected, we investigated the consequences of a position dependent scaling (local conformal transformation) of the metric that destroyed the invariance of the isochoric part too. Motivated by the Weyl geometry, we then introduced the notion of a gauge covariant derivative through a 1-form field and thus obtained a minimally replaced Lagrangian that preserved the invariance of the isochoric part of the Lagrangian under such a local conformal transformation. Using the gauge covariant derivative, we then wrote the modified Lagrange density, which was interpreted as an imposition of the Weyl condition through the Lagrangian. Next, for a non-trivial evolution of the 1-form field, we augmented the Lagrange density with a term involving the exterior derivative of the 1-form field. Having derived the Euler-Lagrange equations through Hamilton's principle, we found a similarity of our equations with governing equations for flexoelectricity under isochoric deformation when we interpreted the exact part of 1-form as the electric field and the anti-exact part as the polarization vector. Similarly, contracting the Weyl condition in various ways, we could model piezoelectricity and electrostriction phenomena. We also looked into magnetomechanical coupling phenomena. We used the Hodge decomposition theorem on the Weylian 1-form that led to the curl of a pseudo-vector field and a vector field. Subsequently, we identified the pseudo-vector field with magnetic potential and the vector part with magnetization and thus modeled flexomagnetism, piezomagnetism and magnetostriction. In summary, starting with the classical elasticity theory and just using a conformal symmetry of the Lagrangian, we could set up a unified framework to interpret and understand several electromechanical and magnetomechanical phenomena.

And yet, we need to do more and nontrivially extend the quasi-static formalism in Roy *et al.* (2019) to the dynamic case so as to systematically understand most electro-magneto-mechanical effects within a unified framework. In order to incorporate time in our formulation, we take a space-time approach where the reference configuration history is defined as a four dimensional (4D) differentiable manifold $\mathcal{B}_r = \mathcal{M}_r \times \mathbb{R}$, i.e. the Cartesian product of a 3-dimensional differentiable manifold $M_r \subset \mathbb{R}^3$ representing the material body with the real line $\mathbb{R}$ denoting the time axis (see also Lagoudas and Edelen, 1989). Thus $\mathcal{B}_r$ is a 4-dimensional cylinder with one dimension parallel to the time axis. In this setting, we define a deformation map whose domain is $\mathcal{B}_r$ and image is another four dimensional manifold $\mathcal{B}_c = \mathcal{M}_c \times \mathbb{R}$ representing the current configuration history. The next step is to prescribe the notion of distance (metric) in the reference and current manifolds. For present purposes, we choose the metric to be of Minkowski type i.e. $\tilde{\mathbf{G}} = \tilde{\mathbf{g}} = \mathrm{diag}\left(1,-1,-1,-1\right)$, where $\tilde{\mathbf{G}}$ and $\tilde{\mathbf{g}}$ are reference and current configuration metrics respectively. With this metric, infinitesimal distances in the reference and current configurations are given as: $dS^2 = \tilde{\mathrm{G}}_{ij} d\tilde{X}^i d\tilde{X}^j$ and $ds^2 = \tilde{\mathrm{g}}_{ij} d\tilde{x}^i d\tilde{x}^j$. Note that we are dealing with a reference configuration in the flat Riemannian space and the field variables (e.g. the displacement field) are defined on the reference domain. These field variables are functions of space-time coordinates assigned to the reference configuration history. If one wishes to measure the distance between two infinitesimally separated points in the current configuration using the coordinates assigned to the reference configuration, then the following equation can be used: $\left(ds\right)^2 = \tilde{\mathbf{C}} d\tilde{\boldsymbol{X}} \cdot d\tilde{\boldsymbol{X}}$ with $\tilde{\mathbf{C}} = \tilde{\mathbf{F}}^T \tilde{\mathbf{g}} \tilde{\mathbf{F}}$ being the 4D right Cauchy-Green tensor (the pulled back metric) and $\tilde{\mathbf{g}}$ the metric in the current configuration. In this space-time setup, the deformation gradient $\tilde{\mathbf{F}}$ and pulled back metric $\tilde{\mathbf{C}}$ both are $4 \times 4$ matrices when represented in terms of 4-coordinates. An interesting observation here is that, departing from classical continuum mechanics, $\tilde{\mathbf{C}}$ contains coupled terms of 3D deformation gradient $\mathbf{F}$ and the material velocity $\mathbf{v}$. After defining the kinematics through $\tilde{\mathbf{C}}$, we need to define an action functional via a Lagrange density function, $\mathcal{L}$. The components of $\tilde{\mathbf{C}}$ include kinetic energy density, the right Cauchy Green tensor $\mathbf{C}$ and terms that involve $\mathbf{F}$ and $\mathbf{v}$ coupling. We construct $\mathcal{L}$ using the kinetic energy density, the invariants of $\mathbf{C}$ and $\mathbf{F}^T\mathbf{v}$ coupled terms with appropriate constants which indicate the material parameters of the system. One may recover the Lagrange density of classical continuum elastodynamics by switching off the coupling constant corresponding to terms involving $\mathbf{F}^T\mathbf{v}$. However, the significance of these terms may be understood by observing the equation of motion. We show that these terms pertain to pseudo-forces in the Euler-Lagrange equations as momentum is modified by various higher order terms. This may have significance in extreme loading conditions where we may need to use the notion of an effective mass. It is widely reported that nonlocal material response may be modelled through the gradient of strain. To account for such effects, one may add a term in the Lagrange density as a function of the covariant derivative of $\tilde{\mathbf{C}}$, which without a loss of generality may be considered isochoric.

After setting up the metric and kinematics, we specifically focus on the important role conformal symmetry plays in solid mechanics. It is well known that hyperelastic materials typically involve energy densities arising from volumetric and shape changes in the material body. If we scale our pulled back metric $\tilde{\mathbf{C}}$ by a constant factor at all space-time points, the volumetric part of the Lagrange density changes and the isochoric part remains unaffected. However, under a localized transformation, i.e. when the scaling factor is a function of position and time in the reference configuration history, the isochoric part is also affected. If one wishes to preserve the symmetry of the Lagrange density under local conformal transformation, then following gauge theory of solids (see Lagoudas and Edelen, 1989), the definition of covariant derivative must be replaced by a gauge covariant derivative. This is accomplished by introducing a new 1-form field $\tilde{\gamma}$ and a specific transformation of that field under scale transformation of the metric. The process of modifying the definition of derivative as $\tilde{\tilde{\nabla}} = \tilde{\nabla} - \tilde{\gamma}$ is referred to as minimal replacement and this in turn weakly imposes a Weyl-like condition in the Lagrange density. However, in order to evolve $\tilde{\gamma}$, we need to add gauge invariant terms associated with the exterior derivative of $\tilde{\gamma}$ in the Lagrange density. With this, we are able to construct a conformal gauge theory which preserves the invariance of the isochoric part of the Lagrangian. We also note that $\tilde{\gamma}$ can be decomposed into an exact, a coexact and an anti-exact part using Hodge decomposition theorem as: $\tilde{\gamma} = \underbrace{\bar{d}\alpha}_{\text{exact}} + \underbrace{\bar{\delta}\boldsymbol{\beta}}_{\text{co-exact}} + \underbrace{\tilde{\tilde{\gamma}}}_{\text{anti-exact}}$ where, $\alpha$, $\boldsymbol{\beta}$ and $\tilde{\tilde{\gamma}}$ are a 0-form (a real-valued smooth function), a 2-form and a 1-form respectively. Hamilton's principle is used to obtain the Euler-Lagrange equations for the displacement field $\mathbf{u}$ and for $\alpha$, $\boldsymbol{\beta}$ and $\tilde{\tilde{\gamma}}$ which may be solved uniquely after specifying appropriate boundary conditions.

As shown in our previous work (see Roy *et al.*, 2019), conformal gauge theory of solids could explain several electromechanical and magnetomechanical phenomena such as piezoelectricity, flexoelectricity, electrostriction, piezomagnetism, flexomagnetism and magnetostriction under quasi-static conditions. There, considering three dimensional (3D) situations (i.e. no time component), we interpreted $\alpha$ as the electric potential, a part of $\boldsymbol{\beta}$ as the magnetic potential and $\bar{\gamma}$ as electric or magnetic polarization as applicable to different cases. However, as we intend to demonstrate these effects in dynamic conditions, we must first confirm that Maxwell's equations for vacuum are recoverable in this framework. We later observe that our conformal gauge theory setup departs from the conventional ideas of electromagnetism, where $\tilde{\gamma}$ is interpreted as the electromagnetic 4-potential which contains the electric scalar potential and the magnetic vector potential. If we also were to use this interpretation, then the electric and magnetic potentials will be directly coupled with $\mathbf{C}$ and $\nabla\mathbf{C}$. But this is clearly unphysical, as we know that only electric and magnetic fields affect strain and not the magnitudes of the associated potentials. Therefore, we need to derive Maxwell's equations from a different perspective. In order to do so, we first note that the electric or magnetic polarization $\bar{\gamma} = \mathbf{0}$ for vacuum and that the number of independent field variables in $\tilde{\gamma}$ is 7. In contrast, in the classical electromagnetic 4-potential, we only have 4 independent variables. Therefore, we postulate that $\boldsymbol{\beta}$ is of the self-dual type. This restriction on $\boldsymbol{\beta}$ leaves only 3 independent variables in it. Writing $\tilde{\gamma}$ explicitly, electric and magnetic fields are then identified. A Lorentz invariant gauge fixing condition is also utilized at this stage so that the classical results are recovered. We realize that if we were to write an energy density term that is quadratic in $\boldsymbol{\gamma}$ to recover the classical results, the system would lose time reversal symmetry and thus become dissipative. Accordingly, we choose terms of the form $\langle \tilde{\gamma}(t), \tilde{\gamma}(-t) \rangle$ in the Lagrange density thereby maintaining the conservative nature of the system. With this, we show that, the energy density for homogeneous Maxwell's equations can be recovered. Finally, the combined Cauchy-Maxwell Lagrangian is written out to lay a foundation for many possible future investigations of a fundamental nature. Some of these might include studies on how pseudo-forces influence the response and dispersive wave propagation in electro-magneto-mechanical materials. As a limited illustration of the working of the space-time theory, a semi-analytical analysis is carried out on an infinite body subjected to isochoric deformation. We also investigate the response of a finite dimensional membrane exhibiting piezoelectricity and piezomagnetism, subjected to both tensile and transverse loading. These exercises reveal that under specific loading frequencies and tension in the membrane, electric and magnetic potentials could increase manifold indicating a possibility of energy harvesting more efficient than hitherto achieved.

The rest of this paper is organized as follows. We present a brief recap of the conformal gauge theory of solids in Section 2 which includes minimal replacement and minimal coupling constructs. A space-time approach is developed to reconcile with classical continuum elastodynamics in Section 3. Following this, a space-time conformal gauge theory of solids is furnished in Section 4. Section 5 deals with the recovery of Maxwell's equations for vacuum based on our theory. We provide illustrative numerical examples considering piezoelectricity and piezomagnetism in Section 6. Finally, a few concluding remarks are presented in Section 7.

## 2 Brief recap of conformal gauge theory of solids

In this section, we present a brief review of conformal gauge theory of solids in line with Roy *et al*. (2019). First, conformal symmetries in classical elasticity are discussed. The concept of local symmetries and minimal replacement are presented next, which is followed up by a derivation of minimal coupling for the Lagrangian. Also included is a physical interpretation of the additional field variables that afford some insight into electromechanical and magnetomechanical coupling phenomena.

## 2.1 Conformal symmetries in classical elasticity

Distance between two infinitesimally close points in the current configuration may be written using the reference configurations coordinates through the following expression.

$$(ds)^2 = g_{\mu\nu} dx^{\mu} dx^{\nu} = \left(\mathbf{F}^{\mathbf{T}}\mathbf{g}\mathbf{F}\right)_{\mu\nu} dX^{\mu} dX^{\nu} = C_{\mu\nu} dX^{\mu} dX^{\nu} \tag{1}$$

Here, $ds$ is the infinitesimal distance in the reference configuration, $\mathbf{F}$ the deformation gradient, $\mathbf{G}$ and $\mathbf{g}$ metrics in the reference and current configurations respectively and $\mathbf{C}$ the right Cauchy-Green tensor. Considering the geometry of the reference configuration to be Euclidean, we see that the geometry described by the metric $\mathbf{C}$ need not be Euclidean. One may scale $\mathbf{C}$ uniformly over the entire body by a constant factor; an operation referred to as conformal transformation and is defined by:

$$\grave{\mathbf{C}} = e^{f} \mathbf{C} \tag{2}$$

where $f$ is a real constant. We now investigate how the Lagrange density is affected by this conformal transformation. As is well known, strain energy density function $\Psi$ can be split into its volumetric ($\Psi_{vol}$) and isochoric ($\Psi_{ic}$) constituents as:

$$\Psi = \Psi_{vol} + \Psi_{ic} \tag{3}$$

Note that, for static deformation, the Lagrange density may be expressed as $\mathcal{L} = -\Psi$. We also know that the materials response often depends on the gradient of strain. In view of this, we choose the following $\Psi_{vol}$ and $\Psi_{ic}$.

$$\Psi_{vol} = c_1 \left[ (\det \mathbf{C})^{\frac{1}{2}} - 1 \right]^2 \tag{4}$$

$$\Psi_{ic} = c_2 C_{\alpha\beta} G^{\alpha\mu} G^{\beta\nu} C_{\mu\nu} (\det \mathbf{C})^{-\frac{2}{3}} + c_3 \nabla_{\alpha} C_{\mu\nu} G^{\alpha\beta} G^{\mu\gamma} G^{\nu\kappa} \nabla_{\beta} C_{\gamma\kappa} (\det \mathbf{C})^{-\frac{2}{3}} \tag{5}$$

Here $\nabla$ denotes the covariant derivative. One may verify that only $\Psi_{vol}$ is affected by the conformal transformation of $\mathbf{C}$ while $\Psi_{ic}$ remains unaffected. Therefore, conformal symmetry exists only in the isochoric part of the classical Lagrange density.

### 2.2 Local symmetry and minimal replacement

If the scale factor is made to depend on position, i.e. $f = f\left(\mathbf{X}\right)$, then under the transformation $\grave{\mathbf{C}} = e^{f(\mathbf{X})}\mathbf{C}$, the invariance of $\Psi_{ic}$ is also lost. In the conformal gauge theory of solids, we preserve the invariance of $\Psi_{ic}$ under local conformal transformations of $\mathbf{C}$ whilst exploiting certain concepts from Weyl geometry. In order to combine gravitation and electricity in a unified framework, Weyl assumed a new geometry of the space-time continuum (Weyl, 1918), where the covariant derivative of the metric tensor was given by a new metric compatibility condition: $\nabla_\alpha g_{\mu\nu} = \gamma_\alpha g_{\mu\nu}$ so that $\gamma$ was an additional 1-form field. Of interest here is the fact that the last condition remains invariant under a group of transformations given by $\grave{\mathbf{g}} = e^f \mathbf{g}$ and $\grave{\gamma} = \gamma + df$. Following Weyl's condition, we modify the definition of the covariant derivative $\nabla$ to a gauge covariant derivative $\overline{\nabla}$ as: $\overline{\nabla} = \nabla - \gamma$. Using Weyl's transformations, i.e. $\grave{\mathbf{C}} = e^{f(\mathbf{X})}\mathbf{C}$ and $\grave{\gamma} = \gamma + df$, one may verify that $\grave{\Psi}_{vol} \neq \Psi_{vol}$ and $\grave{\Psi}_{ic} = \Psi_{ic}$. Thus the invariance of $\Psi_{ic}$ is restored.

### 2.3 Minimal coupling

In order to evolve the 1-form field, we need to add to the Lagrangian gauge invariant terms based on curvature associated with the connection 1-forms. This operation is known as minimal coupling. $\gamma$ may be decomposed into an exact ($\gamma_e$) and an anti-exact ($\overline{\gamma}$) part as follows:

$$\gamma = \gamma_e + \overline{\gamma} = d\lambda + \overline{\gamma} \tag{6}$$

Here, $\lambda$ is a 0-form field whose exterior derivative is the exact part of $\gamma$. The gauge invariant quantity may be written taking the exterior derivative of $\gamma$:

$$d\gamma = d^2\lambda + d\bar{\gamma} = d\bar{\gamma} = \mathbf{Z} \tag{7}$$

Now, the minimal coupling term in the Lagrange density is given as:

$$\Psi_{MC} = \frac{c_4}{4} Z_{ik} G^{ip} G^{kq} Z_{pq} \tag{8}$$

Thus the minimally coupled $\Psi$ may be written as:

$$\begin{aligned}\Psi = {} & c_1\left[\left(\det\mathbf{C}\right)^{\frac{1}{2}} - 1\right]^2 + c_2 \mathrm{C}_{\alpha\beta} G^{\alpha\mu} G^{\beta\nu} \mathrm{C}_{\mu\nu} \left(\det\mathbf{C}\right)^{-\frac{2}{3}} \\ & + c_3\left[\nabla_\alpha \mathrm{C}_{\mu\nu} - \left(d\lambda + \bar{\gamma}\right)_\alpha \mathrm{C}_{\mu\nu}\right] G^{\alpha\beta} G^{\mu\gamma} G^{\nu\kappa} \left[\nabla_\beta \mathrm{C}_{\gamma\kappa} - \left(d\lambda + \bar{\gamma}\right)_\beta \mathrm{C}_{\gamma\kappa}\right] \left(\det\mathbf{C}\right)^{-\frac{2}{3}} \\ & + \frac{c_4}{4} Z_{ik} G^{ip} G^{kq} Z_{pq}\end{aligned} \tag{9}$$

## 2.4 Interpretations of additional field variables

Note that three independent field variables describe the model kinematics, viz. the displacement $\mathbf{u}$, the exact ($\lambda$) and the anti-exact ($\bar{\gamma}$) parts of the 1-form field $\gamma$. In order to find the nature of physical phenomena which may be described by the new Lagrange density, one observes that $\lambda$ and $\bar{\gamma}$ are coupled with $\mathbf{C}$ and $\nabla\mathbf{C}$ in the Lagrangian. Thus whenever strain or strain gradients appear in the material body, an unknown scalar field $\lambda$ and a 1-form field $\bar{\gamma}$ also appear. This gave us motivation to think that the conformal gauge theory might explain electromechanical coupling phenomena if we interpreted the exact part of the 1-form as the electric field and the anti-exact part as the polarization vector. Subsequently, we modeled various electromechanical coupling phenomena, e.g. piezoelectricity, flexoelectricity and electrostriction by contracting the Weyl condition in various ways in the Lagrange density. We also modeled magnetomechanical phenomena, e.g. flexomagnetism, piezomagnetism and magnetostriction using the Hodge decomposition theorem on the 1-form which led to the curl of a pseudo-vector field and a vector field. We identified the pseudo-vector with magnetic potential and the vector with magnetization.

## 3. A space-time formulation of classical continuum mechanics

The stage is now set to consider a reformulation of the classical continuum mechanics in space-time setup. First, we describe the four dimensional reference and current configurations. Having defined the kinematics and derived the pulled back metric in this setup, we describe a construction procedure for the Lagrange density. We also discuss how the elastodynamic Lagrange density is recoverable from our formulation.

### 3.1 Kinematics

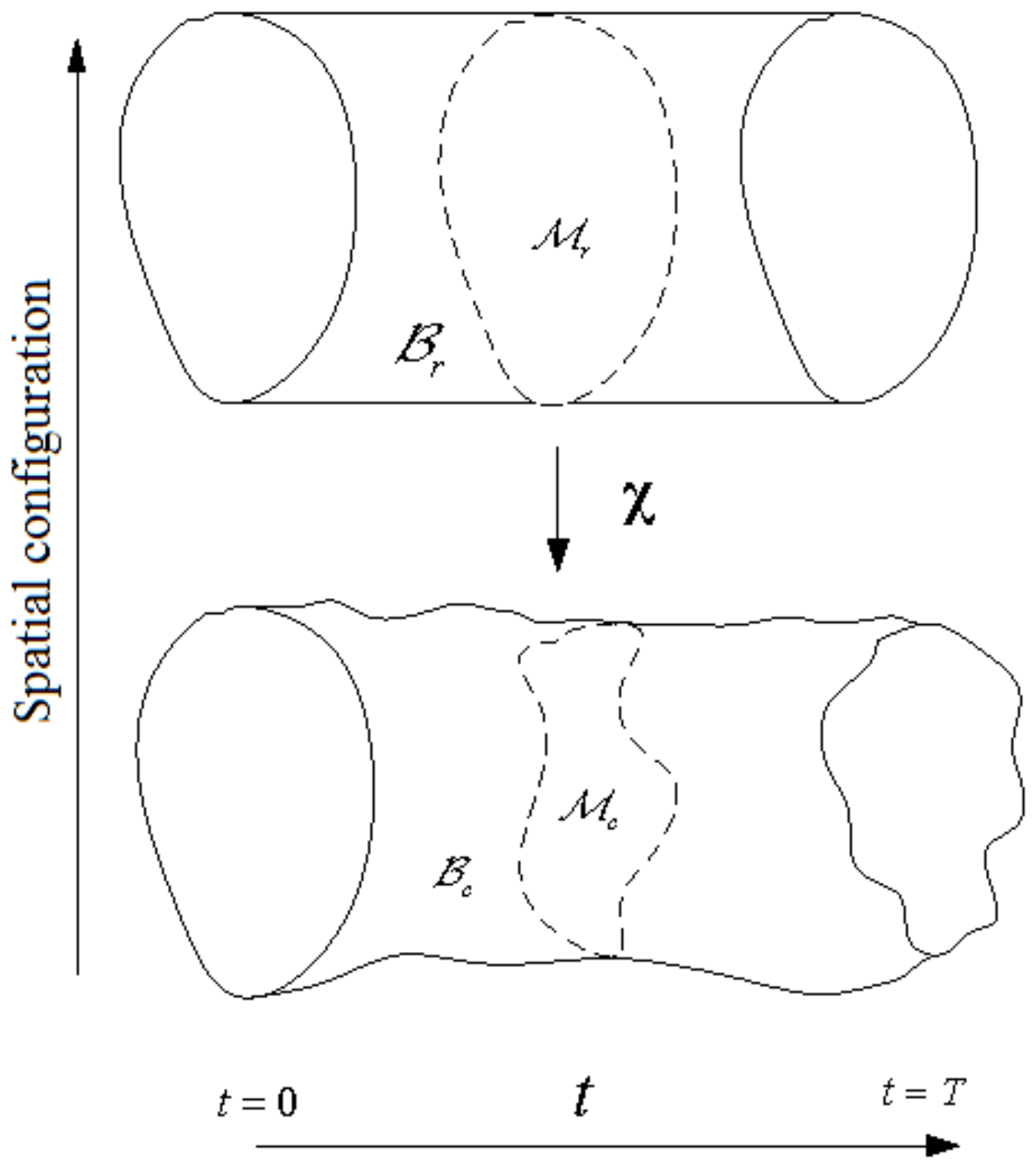

**Figure 1:** Reference and current configuration histories

In order to include time as a fourth coordinate in our theory, we first define the reference and current configurations histories. The reference configuration history is a collection of self-similar copies of the reference configuration at every time point. On the other hand, the current configurations history is a set which includes one reference configuration at time $t=0$ and all the current configurations as the time passes. Mathematically, both are treated as four dimensional (4D) differentiable manifolds, viz. the reference configurations history $\mathcal{B}_r = \left\{ \mathcal{M}_r \big|_t \right\}_{t \in \mathbb{R}}$ and the current configurations history $\mathcal{B}_c = \left\{ \mathcal{M}_c \big|_t \right\}_{t \in \mathbb{R}}$ (see Figure 1) where $M_r \subset \mathbb{R}^3$ and $M_c \subset \mathbb{R}^3$ represent reference and current configurations at some time $0 \le t \in \mathbb{R}$. Time coordinate is along the axis of the cylindrical configuration as shown in Figure 1 and cross sections through the reference and current configuration histories at any time $t$ represent the reference and current configurations of the body at that time. Time coordinate is the same for both reference and current configurations, i.e. deformation of time is not considered here.

Using the Minkowski metric, distances between two infinitesimally close points in $\mathcal{B}_r$ and $\mathcal{B}_c$ may be written using the following expressions.

$$\left(dS\right)^2 = c_{el}^2 \left(dt\right)^2 - \left(d\tilde{X}^2\right)^2 - \left(d\tilde{X}^3\right)^2 - \left(d\tilde{X}^4\right)^2 \tag{10}$$

$$\left(ds\right)^2 = c_{el}^2 \left(dt\right)^2 - \left(d\tilde{x}^2\right)^2 - \left(d\tilde{x}^3\right)^2 - \left(d\tilde{x}^4\right)^2 \tag{11}$$

where, $c_{el}$ is a material parameter which has the units of velocity. It is introduced to ensure consistency among units of space and time coordinates. Equations (10) and (11) may recast in a compact form as follows:

$$dS^2 = \tilde{\mathrm{G}}_{ij} d\tilde{X}^i d\tilde{X}^j \tag{12}$$

$$ds^2 = \tilde{\mathrm{g}}_{ij} d\tilde{x}^i d\tilde{x}^j \tag{13}$$

We use overhead tilde $\left(\tilde{\ }\right)$ to distinguish 4D objects from 3D and indices on a 4D object vary from 1 to 4. $\mathbf{X} = \left(\tilde{X}^2\ \ \tilde{X}^3\ \ \tilde{X}^4\right)^T$ and $\mathbf{x} = \left(\tilde{x}^2\ \ \tilde{x}^3\ \ \tilde{x}^4\right)^T$ denote space coordinates of $\mathcal{B}_r$ and $\mathcal{B}_c$ respectively. We have $\tilde{x}^1 = \tilde{X}^1 = c_{el}\, t = \hat{t}$, where $\hat{t}$ has the unit of length which is the scaled time coordinate of both $\mathcal{B}_r$ and $\mathcal{B}_c$. $\tilde{\mathbf{G}}$ and $\tilde{\mathbf{g}}$ respectively denote Minkowski metric tensors of $\mathcal{B}_r$ and $\mathcal{B}_c$. They are given by $\tilde{\mathbf{G}} = \tilde{\mathbf{g}} = \begin{bmatrix} 1 & \mathbf{0}^T \\ \mathbf{0} & -\mathbf{I} \end{bmatrix}$ with $\mathbf{0} = (0\ 0\ 0)^T$ and $\mathbf{I}$ the $3\times 3$ identity matrix. 4D coordinate differentials in reference and current configurations are given by:

$$d\tilde{X} = \left(c_{el}\, dt\ d\tilde{X}^2\ d\tilde{X}^3\ d\tilde{X}^4\right)^T \tag{14}$$

$$d\tilde{x} = \left(c_{el}\, dt\ d\tilde{x}^2\ d\tilde{x}^3\ d\tilde{x}^4\right)^T \tag{15}$$

Now we define a 4D deformation map that relates the space-time reference and current configurations as:

$$\tilde{x}^i = \tilde{\boldsymbol{\chi}}^i\left(\mathbf{X}, t\right) = \hat{\boldsymbol{\chi}}^i\left(\mathbf{X}, \hat{t}\right) \tag{16}$$

Recall that $i$ varies from 1 to 4. $\tilde{\boldsymbol{\chi}}$ and $\hat{\boldsymbol{\chi}}$ respectively denote deformation maps with real time and scaled time. The coordinate differentials in the current configuration history may be expressed as:

$$d\tilde{x}^1 = d\tilde{X}^1 \tag{17}$$

$$d\tilde{x}^{\alpha+1} = d\mathrm{x}^{\alpha} = \frac{\partial \hat{\chi}^{\alpha+1}}{\partial \mathrm{X}^{\beta}} d\mathrm{X}^{\beta} + \frac{\partial \hat{\chi}^{\alpha+1}}{\partial \hat{t}} d\hat{t} = \mathrm{F}_{\alpha\beta} d\mathrm{X}^{\beta} + \hat{\mathrm{v}}_{\alpha} d\hat{t} \tag{18}$$

where, $\alpha, \beta \in (1,2,3)$. $\mathbf{F}$ is the deformation gradient and $\hat{\mathrm{v}}_{\alpha} = \frac{\partial \hat{\chi}^{\alpha}}{\partial \hat{t}} = \frac{1}{c_{el}} \mathrm{v}_{\alpha}$ the scaled material velocity; $\mathrm{v}_{\alpha}$ is the material velocity. Temporal derivative of the deformation map in equation (18) does not appear in classical elasticity theory, wherein the deformed configuration is assumed as a snapshot of $\mathcal{B}_c$ (space-time current configuration history) at a time *t*. Unlike our space-time formulation, time is not a coordinate in the classical elasticity formulation. As we shall see, by not considering time as coordinate, various types of inertial effects are missed. We may now arrange equations (17) and (18) in a matrix-vector form as:

$$\begin{pmatrix} d\hat{t} \\ d\mathbf{x} \end{pmatrix} = \begin{bmatrix} 1 & \mathbf{0}^T \\ \hat{\mathbf{v}} & \mathbf{F} \end{bmatrix} \begin{pmatrix} d\hat{t} \\ d\mathbf{X} \end{pmatrix} \tag{19}$$

where, $\hat{\mathbf{v}} = \left( \hat{\mathrm{v}}_1 \ \hat{\mathrm{v}}_2 \ \hat{\mathrm{v}}_3 \right)^T$. Using equation (19), equation (13) may be recast as:

$$(ds)^2 = \left( d\hat{t} \ \ d\mathbf{x} \right) \begin{bmatrix} 1 & \mathbf{0}^T \\ \mathbf{0} & -\mathbf{I} \end{bmatrix} \begin{pmatrix} d\hat{t} \\ d\mathbf{x} \end{pmatrix} = \left( \begin{bmatrix} 1 - \hat{\mathbf{v}}^T \hat{\mathbf{v}} & -\hat{\mathbf{v}}^T \mathbf{F} \\ -\mathbf{F}^T \hat{\mathbf{v}} & -\mathbf{F}^T \mathbf{F} \end{bmatrix} \begin{pmatrix} d\hat{t} \\ d\mathbf{X} \end{pmatrix} \right)^T \begin{pmatrix} d\hat{t} \\ d\mathbf{X} \end{pmatrix} = \tilde{C}_{ij} d\tilde{X}^i d\tilde{X}^j \tag{20}$$

Note that in equation (20), we wish to find the distance between two infinitesimally separated points in $\mathscr{B}_c$ in terms of the coordinates of $\mathscr{B}_r$. The 4D pulled back metric $\tilde{\mathbf{C}}$ is given by the following expression.

$$\tilde{\mathbf{C}} = \begin{bmatrix} 1 - \mathbf{v}^T \mathbf{v} / c_{el}^2 & -\mathbf{v}^T \mathbf{F} / c_{el} \\ -\mathbf{F}^T \mathbf{v} / c_{el} & -\mathbf{C} \end{bmatrix} \tag{21}$$

where $\mathbf{C} = \mathbf{F}^T \mathbf{F}$ is the right classical Cauchy-Green tensor. A relation between the 4D pulled back metric and the 3D right Cauchy-Green tensor is obtainable as,

$$\det\left(\tilde{\mathbf{C}}\right) = -\det\left(\mathbf{C}\right) \tag{22}$$

### 3.2 Lagrange density functional

After deriving the 4D pulled back metric $\tilde{\mathbf{C}}$, we try to construct the Lagrange density. In classical continuum mechanics, the Lagrange density is defined as the difference between kinetic energy and potential energy densities. While kinetic energy is a function of the material velocities, potential energy is typically constructed through the invariants of the right Cauchy-Green tensor, $\mathbf{C}$ e.g. $I_1(\text{trace})$, $I_2$ and $I_3(\text{determinant})$. Note from equation (21) that the $(1,1)$ component of 4D pulled back metric tensor is the scaled kinetic energy term. Upon examining various components of the 4D pulled back metric, it is apparent that a coupling between deformation gradient ($\mathbf{F}$) and material velocity ($\mathbf{v}$) naturally exists in our theory naturally – a point of departure from classical continuum mechanics. The reason for this coupling in our formulation lies in the geometric descriptions of the 4D reference and current configuration histories and the map that relates them. If we choose to deal with a conservative system, then the term corresponding to $\mathbf{F}^T\mathbf{v}$ should appear in the Lagrange density in such way that it maintains the time reversal symmetry of the Lagrangian; in other words, $\mathbf{v}$ should appear even number of times in a term. One may also speculate that this new $\mathbf{Fv}$ coupling may bring in nontrivial modifications in the momentum and hence various pseudo-forces that cannot be seen from a Galilean reference frame.

Keeping all these considerations in mind, we construct the invariant Lagrange density under rigid body rotation and translation using various components of the 4D pulled back metric, $\tilde{\mathbf{C}}$ and the invariants of right Cauchy-Green tensor, $\mathbf{C}$ in the following manner. The invariant Lagrange density can be written as,

$$\mathcal{L} = -\Psi \tag{23}$$

where the energy density, $\Psi$ is the sum of different components of the energy densities given as follows,

$$\Psi = \Psi_k + \Psi_{vol} + \Psi_{ic} + \Psi_{add} \tag{24}$$

Here the first term in the energy density is given as,

$$\Psi_k = k_1\left(\tilde{\mathrm{C}}_{11} - 1\right) = -k_1\frac{\mathbf{v}^T\mathbf{v}}{c_{el}^2} \tag{25}$$

The isochoric and volumetric parts of the energy densities are given below in equations (26) and (27) respectively.

$$\Psi_{ic} = k_3 \mathrm{C}_{\mu\nu}\mathrm{G}^{\mu\rho}\mathrm{G}^{\nu\sigma}\mathrm{C}_{\rho\sigma}\left(-\det\tilde{\mathbf{C}}\right)^{-\frac{2}{3}} \tag{26}$$

$$\Psi_{vol} = k_2\left[\left(-\det\tilde{\mathbf{C}}\right)^{\frac{1}{2}} - 1\right]^2 \tag{27}$$

The additional energy density from $\mathbf{F}$ and $\mathbf{v}$ coupled terms is written as:

$$\Psi_{add} = \frac{k_4}{c_{el}^2}\left[\left(\mathbf{F}^{\mathrm{T}}\mathbf{v}\right)\cdot\left(\mathbf{F}^{\mathrm{T}}\mathbf{v}\right)\right] + \frac{k_5}{c_{el}^2}\left[\left(\mathbf{F}:\mathbf{F}\right)\left(\mathbf{v}\cdot\mathbf{v}\right)\right] \tag{28}$$

These coupling terms are not present in the Lagrange density for classical elasticity. A consequence of it is that, the momentum is modified in the Euler-Lagrange equations. As already noted, these modification are of interest as they lead to various pseudo-forces in the Euler-Lagrange equation.

### 3.3 Further insights and recovering the classical Lagrange density

We now discuss how the momentum in our theory is modified with additional terms. A quick appraisal of these terms offer certain insights into the nature of the pseudo-forces. We also show how the classically known elasto-dynamic Lagrange density is recoverable from our theory.

#### 3.3.1 Modifications in the momentum

We use the Lagrange density in equation (23) to obtain the corresponding Euler-Lagrange equation. To understand the implications of the terms with $\mathbf{F}, \mathbf{v}$ coupling, we may write the relevant terms of the momentum as follows:

$$\mathrm{M}_k = \frac{2k_4}{c_{el}^2}\mathrm{F}_{mi}\mathrm{v}_m\mathrm{F}_{ki} + \frac{2k_5}{c_{el}^2}\mathrm{v}_k\mathrm{F}_{ij}\mathrm{F}_{ij} \tag{29}$$

It is evident that the deformation gradient plays a significant role in these terms, i.e. the effective mass is dependent on the deformation gradient. If the deformation gradient is high, these terms will influence and alter the dynamics in ways that are beyond the predictive realm of classical elastodynamics. We may carry out a simple analysis to draw some insight from these terms. Considering a rigid body motion i.e. $\mathbf{x} = \mathbf{Q}(t)\mathbf{X} + \mathbf{c}(t)$, we may evaluate the momentum presented in equation (29) as follows:

$$\mathbf{M} = \frac{2}{c_{el}^2}\left(k_4 + 3k_5\right)\mathrm{v}_k \tag{30}$$

Clearly, the resulting inertia terms will contain the acceleration $\dot{\mathbf{v}}$. For rigid body motion, we may express the velocity as $\mathbf{v} = \boldsymbol{\omega}\times\left(\mathbf{x}-\mathbf{c}\right)+\dot{\mathbf{c}}$ and the acceleration as $\dot{\mathbf{v}} = \dot{\boldsymbol{\omega}}\times\left(\mathbf{x}-\mathbf{c}\right)+\frac{1}{2}\boldsymbol{\omega}\times\left(\boldsymbol{\omega}\times\left(\mathbf{x}-\mathbf{c}\right)\right)+\frac{1}{2}\boldsymbol{\omega}\times\left(\mathbf{v}-\dot{\mathbf{c}}\right)+\ddot{\mathbf{c}}$ where $\boldsymbol{\omega}$ is the axial vector of $\dot{\mathbf{Q}}\mathbf{Q}^T$. Therefore, it is evident that centrifugal $\left(\boldsymbol{\omega}\times\boldsymbol{\omega}\times\mathbf{x}\right)$, Coriolis $\left(\boldsymbol{\omega}\times\mathbf{v}\right)$, and Euler $\left(\dot{\boldsymbol{\omega}}\times\mathbf{x}\right)$ type pseudo-forces naturally appear in the equation of motion. It is understood that while writing the Lagrangian, we have not made *a priori* any assumption about the type of the reference frame (e.g. inertial or non-inertial). In the presence of non-rigid motion, deformation gradient will modify these forces in a more involved way. Among other scenarios, these forces might have significance under extreme loading conditions.

### 3.3.2 Recovery of elastic Lagrange density

The classical Lagrange density for elastodynamics is defined as,

$$\mathcal{L}^{elas} = T - V \tag{31}$$

where, $T$ and $V$ are respectively the kinetic energy and potential energy densities given by

$$T = \frac{1}{2}\rho\left(\mathrm{v}^T\mathrm{v}\right)$$

$$V = \frac{\lambda}{2}\left(\mathrm{tr}(\boldsymbol{\varepsilon})\right)^2 + \mu\,\boldsymbol{\varepsilon}:\boldsymbol{\varepsilon} \tag{32}$$

Infinitesimal strain is $\boldsymbol{\varepsilon} \triangleq \frac{1}{2}\left(\nabla\mathrm{u}+\nabla\mathrm{u}^T\right)$, where $\mathbf{u}$ is the displacement vector. The deviatoric strain, $\boldsymbol{\varepsilon}^d$ is given as, $\boldsymbol{\varepsilon}^d = \boldsymbol{\varepsilon} - \frac{\mathrm{tr}\left(\boldsymbol{\varepsilon}\right)}{n}\mathbf{I}$, where $\mathrm{tr}\left(.\right)$ denotes the trace of a matrix or a tensor (as appropriate). $\lambda$ and $\mu$ are Lamé's parameters and $n$ is the spatial dimension. Comparing the classical Lagrange density in Eq. (31) with our space-time Lagrange density in Eq. (23), we see that $\Psi_k$ contains the kinetic energy density if we set $k_1 = 0.5\rho c_{el}^2$.

One may anticipate that the constants for the terms with $\mathbf{F}, \mathbf{v}$ coupling (i.e. $k_4$ and $k_5$) are possibly small in the classical loading regime. Hence we set both coupling constants to zero. For recovering the material constants $k_2$ and $k_3$ corresponding to the volumetric and isochoric parts of energy, we compare the Cauchy stress obtained from the classical elastic Lagrange density with the linearized stress based on our space-time formulation. The Cauchy stress is given as,

$$\sigma = k_n \operatorname{tr}(\boldsymbol{\varepsilon})\mathbf{I} + 2\mu\boldsymbol{\varepsilon}^d \tag{33}$$

where, $k_n = \lambda + \frac{2\mu}{n}$ is the $n$-dimensional bulk modulus. The second Piola-Kirchhoff stress computed from our space-time Lagrange density is,

$$\mathbf{S} := 2\frac{\partial \Psi}{\partial \mathbf{C}} = 2k_2\left[(\det \mathbf{C})^{\frac{1}{2}} - 1\right](\det \mathbf{C})^{\frac{1}{2}}\mathbf{C}^{-1} + 4k_3\left[\mathbf{C} - \frac{1}{3}(\mathbf{C}:\mathbf{C})\mathbf{C}^{-1}\right](\det \mathbf{C})^{-\frac{2}{3}} \tag{34}$$

Now using the approximations, $\mathbf{C} \approx \mathbf{I} + 2\boldsymbol{\varepsilon}$, $\mathbf{C}^{-1} \approx \mathbf{I} - 2\boldsymbol{\varepsilon}$ and $\det(\mathbf{C}) \approx 1 + 2\operatorname{tr}(\varepsilon)$ in equation (34) and setting $k_2 = \frac{1}{2}\left(\lambda + \frac{2\mu}{n}\right)$ and $k_3 = \mu/8$, the stress expressions in equations (33) and (34) match and we thus recover the classical Lagrange density for elastodynamics through our space-time formulation.

## 4 Gauge theory of solids using conformal symmetry

Predicting the behavior of solids exhibiting electromechanical and magnetomechanical response under large deformation and dynamic loading conditions has remained a formidable challenge for over a century. Mathematical models for these problems are typically through an energy functional that considers possible coupling between electric fields and strains (in electromechanical coupling) or magnetic fields and strains (in magnetomechanical cases). These coupling terms are phenomenological and lack a basis in the deep underlying symmetry requirements of the Lagrangian. Extending these models to finite deformation cases is also far from trivial. A much desirable approach in this context would be a unified framework through which one may analyze these phenomena in a rational manner. The gauge theory of solids is such a platform wherein certain local symmetries on the Lagrangian are exploited to arrive at the conservation laws. Lagoudas and Edelen (1989) have considered a gauge theoretic approach to explain the dynamic behavior of solids with (a class of) continuously distributed defects. Exploring the local action of the spatial symmetry group $G_s = SO(3) \triangleright T(3)$ and the material symmetry group $G_m = \{SO(3) \triangleright T(3)\} \times T(1)$, they have tried to explain damage (through microcracking) and plasticity (through dislocation motion) from a fundamental perspective. Recently, Roy *et al*. (2019) have developed a conformal gauge theory of solids to explain electromechanical and magnetomechanical phenomena under static conditions. In this section, we extend the conformal gauge theory to dynamic conditions using the space-time strategy developed in Section 3.

### 4.1 Local conformal transformation of the pulled back metric

One may write the conformal transformation of the pulled back metric $\tilde{\mathbf{C}}$ as:

$$`\tilde{\mathbf{C}} = e^f \tilde{\mathbf{C}} \tag{35}$$

where $f$ is a real constant. Consider the effect of this transformation on the Lagrange density defined in equation (23). Equation (24) may be recast as follows:

$$\Psi = \bar{\Psi} + \Psi_{ic} \tag{36}$$

where, $\bar{\Psi} = \Psi_k + \Psi_{vol} + \Psi_{add}$. Thanks to recent progresses in the non-local mechanics of solids, we know that the response of materials may depend on strain gradients. In order to account for gradient effects, we may add such terms in the Lagrange density, which without a loss of generality may be considered isochoric. The expression of $\Psi_{ic}$ may be written as:

$$\Psi_{ic} = k_3 \mathrm{C}_{\mu\nu} \mathrm{G}^{\mu\rho} \mathrm{G}^{\nu\sigma} \mathrm{C}_{\rho\sigma} \left(-\det \tilde{\mathbf{C}}\right)^{-\frac{2}{3}} + h_{ijkpqr} \tilde{\mathrm{G}}^{ia} \tilde{\mathrm{G}}^{jb} \tilde{\mathrm{G}}^{kc} \tilde{\nabla}_a \tilde{\mathrm{C}}_{bc} \tilde{\mathrm{G}}^{pd} \tilde{\mathrm{G}}^{qe} \tilde{\mathrm{G}}^{rf} \tilde{\nabla}_d \tilde{\mathrm{C}}_{ef} \left(-\det \tilde{\mathbf{C}}\right)^{-\frac{2}{3}} \quad (37)$$

One may verify that, under the conformal transformation of $\tilde{\mathbf{C}}$, the following holds.

$$`\bar{\Psi} \neq \bar{\Psi} \quad (38)$$

$$`\Psi_{ic} = \Psi_{ic} \quad (39)$$

However, under a position and time dependent conformal transformation, $`\tilde{\mathrm{C}} = e^{f(\mathbf{X},t)} \tilde{\mathrm{C}}$, $\bar{\Psi}$ and $\Psi_{ic}$ transform in such a way that,

$$`\bar{\Psi} \neq \bar{\Psi} \quad (40)$$

$$`\Psi_{ic} \neq \Psi_{ic} \quad (41)$$

Therefore, the invariance of $\Psi_{ic}$ is lost. To restore the invariance, we need minimal replacement and hence a gauge covariant definition of derivative.

## 4.2 Minimal replacement using Weyl condition

In order to render $\Psi_{ic}$ invariant, the covariant derivative $\tilde{\nabla}$ is modified to a gauge covariant derivative $\tilde{\tilde{\nabla}}$ as:

$$\tilde{\tilde{\nabla}} = \tilde{\nabla} - \tilde{\boldsymbol{\gamma}} \quad (42)$$

One may observe that under Weyl's transformations on the pulled back metric i.e. $`\tilde{\mathbf{C}} = e^f \tilde{\mathbf{C}}$ and $`\tilde{\boldsymbol{\gamma}} = \tilde{\boldsymbol{\gamma}} + \tilde{d}f$, the following holds.

$$\tilde{\tilde{\nabla}} `\tilde{\mathbf{C}} = \tilde{\nabla} `\tilde{\mathbf{C}} - `\tilde{\boldsymbol{\gamma}} `\tilde{\mathbf{C}} = e^{f(\mathbf{X},t)} \tilde{\tilde{\nabla}} `\tilde{\mathbf{C}} \quad (43)$$

With this modified definition of covariant derivative, $\Psi_{vol}$ and $\Psi_{ic}$ transform as:

$$`\Psi_{vol} \neq \Psi_{vol} \tag{44}$$

$$`\Psi_{ic} = \Psi_{ic} \tag{45}$$

Hence the invariance of $\Psi_{ic}$ is preserved. We may now write the modified energy density for the space-time conformal Lagrange density by replacing $\tilde{\nabla}\tilde{\mathbf{C}}$ with $\tilde{\bar{\nabla}}\tilde{\mathbf{C}}$ in $\Psi_{vol}$ and $\Psi_{ic}$ as follows:

$$\Psi = \bar{\Psi} + \Psi_{ic}\left(\mathbf{C}, \tilde{\bar{\nabla}}\tilde{\mathbf{C}}\right) \tag{46}$$

where,

$$\begin{aligned}\Psi_{ic} &= k_3 \mathrm{C}_{\mu\nu} \mathrm{G}^{\mu\rho} \mathrm{G}^{\nu\sigma} \mathrm{C}_{\rho\sigma} \left(-\det \tilde{\mathbf{C}}\right)^{-\frac{2}{3}} \\ &\quad + h_{ijkpqr} \tilde{\mathrm{G}}^{ia} \tilde{\mathrm{G}}^{jb} \tilde{\mathrm{G}}^{kc} \left(\tilde{\nabla}_a \tilde{\mathrm{C}}_{bc} - \tilde{\gamma}_a \tilde{\mathrm{C}}_{bc}\right) \tilde{\mathrm{G}}^{pd} \tilde{\mathrm{G}}^{qe} \tilde{\mathrm{G}}^{rf} \left(\tilde{\nabla}_d \tilde{\mathrm{C}}_{ef} - \tilde{\gamma}_d \tilde{\mathrm{C}}_{ef}\right)\left(-\det \tilde{\mathbf{C}}\right)^{-\frac{2}{3}}\end{aligned}$$

**4.3 Minimal Coupling**

In order to capture the evolution of the additional field variable $\tilde{\gamma}$, we add gauge invariant terms based on curvature associated with the connection 1-forms to the Lagrangian. One may construct a gauge invariant quantity from $\tilde{\boldsymbol{\gamma}}$ in the following way.

$$\mathbf{Z} = \tilde{d}\tilde{\boldsymbol{\gamma}} \tag{47}$$

Note that $\mathbf{Z}$ is the curvature 2-form. We write the minimal coupling term in the Lagrange density as:

$$\Psi_{MC} = \frac{c_4}{4} Z_{ik} \tilde{G}^{ip} \tilde{G}^{kq} Z_{pq} \tag{48}$$

Finally, we may write the energy density for the combined electromechanical and magnetomechanical cases as follows:

$$\Psi = \bar{\Psi} + \Psi_{ic}\left(\mathbf{C}, \tilde{\bar{\nabla}}\tilde{\mathbf{C}}\right) + \Psi_{MC} \tag{49}$$

Here, $\Psi_{MC}$ is the minimal coupling term constructed through the gauge invariant curvature term using the 1-form field $\tilde{\boldsymbol{\gamma}}$ .

It is evident from equation (49) that electric and magnetic fields also affect the momentum . This is again a departure from the phenomenological theories of electro-magneto-mechanical coupling. It may provide useful hints on controlling the effective mass through the induced electric field, thereby affording a possible framework, say, for the design of materials and structures where crack initiation and propagation could perhaps be controlled. Another interesting study would be to investigate the nature of waves in solid materials which exhibit electro-magneto-mechanical effects. Similar studies have been carried out in magnetohydrodynamics where Navier-Stokes and Maxwell's equations are combined in a single framework. This explains the behavior of electrically conducting fluids under external magnetic fields. One important aspect of this coupled behavior is the presence of Alfven waves in such fluids which arises out of the coupling among Navier-Stokes equation and Maxwell's equations. As our theory bears similarity with magnetohydrodynamics and differs only in the type of material, which is solid in our case, we suspect that one may find Alfven like waves in solid materials by carrying out dispersion analysis.

## 5 Recovering energy density for Maxwell's equations in vacuum

We now demonstrate the recovery of Maxwell's equations from our theory. To remain consistent with the electro-magneto-mechanical coupling in solid materials, we need to deviate from the classical ideas of electromagnetism where $\tilde{\gamma}$ is interpreted as the electromagnetic 4-potential. If we stick to this interpretation, then electric potential and magnetic potentials will be directly coupled with $\mathbf{C}$ and $\nabla\mathbf{C}$ which is unphysical as only electric and magnetic fields affect strain and not the associated potentials. Accordingly, we take a different viewpoint and use the Hodge decomposition of $\tilde{\gamma}$ . Our analysis provides a new derivation of Maxwell's equations.

In our previous work (Roy *et al.*, 2019) that dealt only with static cases, we had incorporated various contracted forms of the Weyl condition in the Lagrangian. Specifically, to model electromechanical coupling phenomena, e.g. piezoelectricity, flexoelectricity, electrostriction, we decomposed the 1-form field as: $\boldsymbol{\gamma} = d\alpha + \bar{\boldsymbol{\gamma}}$ with the interpretation that $\alpha$ and $\bar{\boldsymbol{\gamma}}$ are respectively proportional to electric potential and electric polarization. For magnetomechanical cases, e.g. piezomagnetism, flexomagnetism and magnetostriction, we used a different form of decomposition: $\boldsymbol{\gamma} = \nabla \times \boldsymbol{\tau} + \bar{\boldsymbol{\gamma}}$, where $\boldsymbol{\tau}$ and $\bar{\boldsymbol{\gamma}}$ were proportional to magnetic potential and magnetization respectively. However, in the case of dynamics, we need to use a more general decomposition of the 1-form field $\tilde{\gamma}$ based on the Hodge decomposition theorem, according to which any *p*-form $\tilde{\boldsymbol{\gamma}} \in \Lambda^p(\mathcal{M})$ on a compact orientable smooth manifold $\mathcal{M}$ of dimension $n$, may be written as (Ivancevic and Ivancevic, 2008):

$$\tilde{\gamma} = \underbrace{\tilde{d}\alpha}_{\text{exact}} + \underbrace{\bar{\delta}\boldsymbol{\beta}}_{\text{co-exact}} + \underbrace{\tilde{\bar{\gamma}}}_{\text{anti-exact}} \tag{50}$$

where $\alpha \in \Lambda^{p-1}(\mathcal{M})$, $\boldsymbol{\beta} \in \Lambda^{p+1}(\mathcal{M})$ and $\tilde{\bar{\boldsymbol{\gamma}}} \in \Lambda^p(\mathcal{M})$. $d\alpha$, $\bar{\delta}\boldsymbol{\beta}$ and $\tilde{\bar{\boldsymbol{\gamma}}}$ are respectively the exact, coexact and harmonic parts of $\tilde{\boldsymbol{\gamma}}$. Here $\Lambda^p$ is the space of differential forms of order *p*. The operator $\bar{\delta} : \Lambda^p(\mathcal{M}) \to \Lambda^{p-1}(\mathcal{M})$ is defined as,

$$\bar{\delta} = (-1)^{n(p+1)+1} s * d * \tag{51}$$

where, $*$ is the Hodge star operator with $s = -1$ for the Minkowski metric. As we intend to demonstrate various electro-magneto-mechanical effects under dynamic conditions with this decomposition of the 1-form field $\tilde{\boldsymbol{\gamma}}$, we first need to ascertain whether Maxwell's equations in vacuum are recoverable from this framework. In our case, $n = 4$, $p = 1$ and $s = -1$. We may write:

$$\tilde{\gamma} = \tilde{d}\alpha + *\tilde{d} * \boldsymbol{\beta} + \tilde{\bar{\gamma}} \tag{52}$$

In the component form, equation (52) may be written as,

$$\begin{pmatrix}\tilde{\gamma}_1\\ \tilde{\gamma}_2\\ \tilde{\gamma}_3\\ \tilde{\gamma}_4\end{pmatrix}=\begin{pmatrix}\partial_1\alpha\\ \partial_2\alpha\\ \partial_3\alpha\\ \partial_4\alpha\end{pmatrix}+\begin{pmatrix}\partial_2\beta_{12}+\partial_3\beta_{13}+\partial_4\beta_{14}\\ -\partial_1\beta_{12}-\partial_3\beta_{23}-\partial_4\beta_{24}\\ -\partial_1\beta_{13}+\partial_2\beta_{23}-\partial_4\beta_{34}\\ -\partial_1\beta_{14}+\partial_2\beta_{24}+\partial_3\beta_{34}\end{pmatrix}+\begin{pmatrix}\tilde{\tilde{\gamma}}_1\\ \tilde{\tilde{\gamma}}_2\\ \tilde{\tilde{\gamma}}_3\\ \tilde{\tilde{\gamma}}_4\end{pmatrix} \tag{53}$$

Denoting $\omega_1=-\beta_{12}$, $\omega_2=-\beta_{13}$, $\omega_3=-\beta_{14}$ and $\bar{\omega}_1=-\beta_{34}$, $\bar{\omega}_2=\beta_{24}$, $\bar{\omega}_3=-\beta_{23}$ and considering $\tilde{\tilde{\boldsymbol{\gamma}}}=0$, we may rewrite equation (53) in a more compact form as follows.

$$\tilde{\gamma}=\begin{pmatrix}\dfrac{\partial\alpha}{\partial\hat{t}}-\nabla\cdot\boldsymbol{\omega}\\ \nabla\alpha+\dfrac{\partial\boldsymbol{\omega}}{\partial\hat{t}}+\nabla\times\bar{\boldsymbol{\omega}}\end{pmatrix} \tag{54}$$

Here, $\boldsymbol{\omega}=\left(\omega_1\ \omega_2\ \omega_3\right)^T$ and $\bar{\boldsymbol{\omega}}=\left(\bar{\omega}_1\ \bar{\omega}_2\ \bar{\omega}_3\right)^T$. In describing $\tilde{\boldsymbol{\gamma}}$ via equation (54), we have used seven independent variables which should be related to electric and magnetic potentials. However, there are only 4 independent variables in the classical description of electric and magnetic potentials. From this, it is evident that the $\boldsymbol{\beta}$ matrix has to be of a special kind – involving only 3 independent variables. In order to achieve this, we consider $\boldsymbol{\beta}$ to be self-dual, i.e. $\boldsymbol{\beta}$ satisfies the following condition,

$$\left(\beta_{dual}\right)_{\mu\nu}=\beta^{\mu\nu} \tag{55}$$

where, $\left(\beta_{dual}\right)^{\alpha\beta}=\frac{1}{2}\in^{\alpha\beta\gamma\delta}\beta_{\gamma\delta}$ with $\left(\beta_{dual}\right)_{\mu\nu}=\tilde{g}_{\mu\alpha}\tilde{g}_{\nu\beta}\left(\beta_{dual}\right)^{\alpha\beta}$. This implies that $\boldsymbol{\omega}=\bar{\boldsymbol{\omega}}$, or in other words, $\omega_1=\beta_{34}=\beta_{12}$, $\omega_2=-\beta_{24}=\beta_{13}$, $\omega_3=\beta_{23}=\beta_{14}$. Using equation (55) and the classical definition of electric field $\mathbf{E}=-\nabla\varphi-\frac{\partial\mathbf{A}}{\partial t}$ and magnetic field $\mathbf{B}=\nabla\times\mathbf{A}$ and with the identification $\alpha=k\varphi/c_{el}$ and $\boldsymbol{\omega}=k\mathbf{A}$ ($k$ is a proportionality constant and $c_{el}$ the electromagnetic wave speed), we may recast equation (54) as:

$$\tilde{\gamma}\left(\mathbf{X},\hat{t}\right)=\begin{pmatrix}\dfrac{\partial\alpha}{\partial\hat{t}}-\nabla\cdot\boldsymbol{\omega}\\ k\left(-\mathbf{E}/c_{el}+\mathbf{B}\right)\end{pmatrix} \tag{56}$$

If we were to use a quadratic form of $\tilde{\gamma}(\mathbf{X},\hat{t})$ in the Lagrange density, the system would lose time reversal symmetry and become dissipative. To maintain the conservative nature of the system, we use the term $\langle \tilde{\gamma}(\mathbf{X},t), \tilde{\gamma}(\mathbf{X},-t) \rangle$ in the Lagrangian that has time reversal symmetry. Denoting $\tilde{\gamma}^{+} = \tilde{\gamma}(\mathbf{X},\hat{t})$ and $\tilde{\gamma}^{-} = \tilde{\gamma}(\mathbf{X},-\hat{t})$, we may thus write the invariant Lagrange density under space-time conformal symmetry as,

$$\mathcal{L}^{em} = k_{em}\, \tilde{\gamma}_i^{+}\, \tilde{g}^{ip}\, \tilde{\gamma}_p^{-} \tag{57}$$

Here, $\tilde{\gamma}^{-}$ is defined as:

$$\tilde{\gamma}^{-} = \tilde{\gamma}(\mathbf{X},-\hat{t}) = \begin{pmatrix} -\dfrac{\partial \alpha}{\partial \hat{t}} + \nabla \cdot \boldsymbol{\omega} \\ k\left(-\mathbf{E}/c_{em} - \mathbf{B}\right) \end{pmatrix} \tag{58}$$

Note that in equation (58), we have made use of the following transformation.

$$\boldsymbol{\omega}(\mathbf{X},-\hat{t}) = -\boldsymbol{\omega}(\mathbf{X},\hat{t}) \tag{59}$$

In order to recover the energy density for Maxwell's equations, we need to impose the following gauge condition,

$$\frac{\partial \alpha}{\partial \hat{t}} - \nabla \cdot \boldsymbol{\omega} = 0 \tag{60}$$

Using condition (60) in (57), one may now recover the energy density for Maxwell's equations in vacuum as follows:

$$\mathcal{L}^{em} = k_c \left(\mathbf{E}\cdot\mathbf{E} - c_{el}^2\, \mathbf{B}\cdot\mathbf{B}\right) \tag{61}$$

where, $k_c = k^2\, k_{em}/c_{el}^2$ is a constant. Note that $\mathcal{L}^{em}$ is the Lagrange density whose Euler-Lagrange equations are homogeneous Maxwell's equations.

## 6 Illustrative examples with piezoelectricity and piezomagnetism

We furnish a semi-numerical study on the space-time conformal gauge theory in this section. Considering materials exhibiting piezoelectricity and piezomagnetism, the material parameters of our theory are determined as functions of the classical parameters in the linearized setup. We analyze an infinite body subjected to isochoric deformation and a finite dimensional membrane subjected to both tensile and transverse loading in the frequency domain.

### 6.1 Energy density and Euler-Lagrange equations

The strain energy density for non-centrosymmetric materials exhibiting piezoelectricity and piezomagnetism in the space-time conformal gauge theory is given as,

$$\begin{aligned}\Psi^{pz} &= \frac{1}{2}\boldsymbol{E}_{\mu\nu}\mathbb{C}_{\mu\nu\alpha\beta}\boldsymbol{E}_{\alpha\beta} + e_{\alpha\mu\nu}\left(\nabla_{\alpha}\mathrm{C}_{\mu\nu} - \gamma_{\alpha}^{+}\mathrm{C}_{\mu\nu}\right)\left(\det\mathbf{C}\right)^{-\frac{1}{3}} \\ &\quad - K_{\alpha\beta}\left(\mathrm{C}_{pq}^{-1}\nabla_{\alpha}C_{pq} - 3\gamma_{\alpha}^{+}\right)\left(\mathrm{C}_{rs}^{-1}\nabla_{\beta}C_{rs} - 3\gamma_{\beta}^{+}\right)\end{aligned} \tag{62}$$

where, $\mathbb{C}$ is the fourth order elastic stiffness tensor. $\boldsymbol{e}$ and $\boldsymbol{K}$ are third and second order material constant tensors respectively. Green-Lagrange strain tensor, $\boldsymbol{E}$ is given as $\boldsymbol{E} = \frac{1}{2}\left(\mathbf{C} - \mathbf{I}\right)$. $\boldsymbol{\gamma}^{+}$ and $\boldsymbol{\gamma}^{-}$ may be written in terms of electric and magnetic fields as,

$$\boldsymbol{\gamma}^{+} = k_6\mathbf{E} + k_7\mathbf{B} \tag{63}$$

$$\boldsymbol{\gamma}^{-} = k_6\mathbf{E} - k_7\mathbf{B} \tag{64}$$

Here $k_6$ and $k_7$ are material constants. The first term on the RHS of equation (62) is the elastic strain energy density for finite deformation elasticity. The second and third terms denote space-time conformal gauge invariant electro-magneto-mechanical coupling.

We write the Lagrange density for quasi-static conditions as $\mathcal{L}^{pz} = -\Psi^{pz}$. For a given deformation, the Euler-Lagrange equations for scalar potential $\varphi$ and vector potential $\mathbf{A}$ are given below in Eqs. (65) and (66) respectively.

$$\nabla_{\alpha}\left(D_{\mathbf{E}}\mathcal{L}^{pz}\right)_{\alpha} = 0 \tag{65}$$

$$\frac{\partial}{\partial t}\left(D_{\mathbf{E}}\mathcal{L}^{pz}\right)_{\alpha} - \nabla_{\mu}\left(D_{\mathbf{B}}\mathcal{L}^{pz}\right)_{\nu}\varepsilon_{\nu\mu\alpha} = \mathbf{0} \tag{66}$$

where,

$$\left(D_{\mathbf{E}}\mathcal{L}^{pz}\right)_{\alpha} = e_{\alpha\mu\nu}k_6 C_{\mu\nu}\left(\det \mathbf{C}\right)^{-\frac{1}{3}} - 6k_6\, K_{\alpha\beta}\mathrm{C}_{rs}^{-1}\nabla_{\beta}\mathrm{C}_{rs} + 18\,K_{\alpha\beta}k_6^2\,\mathrm{E}_{\beta} \tag{67}$$

$$\left(D_{\mathbf{B}}\mathcal{L}^{pz}\right)_{\alpha} = e_{\alpha\mu\nu}k_7 C_{\mu\nu}\left(\det \mathbf{C}\right)^{-\frac{1}{3}} - 18\,K_{\alpha\beta}\,k_7^2\,\mathrm{B}_{\beta} \tag{68}$$

Note that, $\boldsymbol{\varepsilon}$ is the alternating tensor in 3D.

### 6.2 Determination of material parameters

We determine the material parameters of our space-time conformal gauge theory by comparing Euler-Lagrange equations derived from the classical piezoelectricity theory with those based on our theory in the linearized setup. Isochoric deformation condition is assumed. We also consider a zero magnetic field in the Euler–Lagrange equations through our theory for correspondence with the classical piezoelectricity equations.

The Lagrange density for classical piezoelectricity may be written as (Tiersten, 2013):

$$\mathcal{L}^{c} = -\left(\frac{1}{2}\overline{\mathbb{C}}_{ijkl}\varepsilon_{ij}\varepsilon_{kl} - \overline{e}_{ijk}\mathrm{E}_{i}\varepsilon_{jk} - \frac{1}{2}\overline{K}_{ij}\mathrm{E}_{i}\mathrm{E}_{j}\right) \tag{69}$$

where, $\overline{\mathbb{C}}$, $\overline{\boldsymbol{e}}$ and $\overline{\boldsymbol{K}}$ are the stiffness tensor, piezoelectric constant tensor and dielectric constant tensor respectively. Considering isochoric deformation and using $\overline{\mathbb{C}}_{ijkl} = \overline{\mathbb{C}}_{klij}$ and $\overline{K}_{ij} = \overline{K}_{ji}$, the variation of the Lagrangian density for classical piezoelectricity may be written as,

$$\delta\mathcal{L}^{c} = -\left[\left(\overline{\mathbb{C}}_{ijkl} - \frac{1}{3}\overline{\mathbb{C}}_{mmkl}\mathrm{I}_{ij}\right)\varepsilon_{kl}^{d} + \left(\overline{e}_{kij} - \frac{1}{3}\overline{e}_{knn}\mathrm{I}_{ij}\right)\mathrm{E}_{k}\right]\delta\varepsilon_{ij} - \overline{K}_{ij}\mathrm{E}_{j}\,\delta\mathrm{E}_{i} \tag{70}$$

The Euler-Lagrange equations are,

$$\nabla\cdot\overline{\sigma} + \mathbf{f} = \mathbf{0} \tag{71}$$

$$\nabla\cdot\overline{\mathbf{D}} = 0 \tag{72}$$

where,

$$\bar{\sigma}_{ij} = \left( \bar{\mathbb{C}}_{ijkl} - \frac{1}{3} \bar{\mathbb{C}}_{mmkl} \mathrm{I}_{ij} \right) \varepsilon^{d}_{kl} - \left( \bar{e}_{nij} - \frac{1}{3} \bar{e}_{nkk} \mathrm{I}_{ij} \right) E_n \tag{73}$$

$$\bar{\mathrm{D}}_i = \bar{e}_{ijk} \varepsilon^{d}_{jk} + \bar{K}_{ij} \mathrm{E}_j \tag{74}$$

Similarly, we can derive the Euler-Lagrange equations from the stain energy density given in equation (62) for isochoric deformation under a zero magnetic field. Using the approximations $\mathbf{C} \approx \mathbf{I} + 2\boldsymbol{\varepsilon}$ , $\mathbf{C}^{-1} \approx \mathbf{I} - 2\boldsymbol{\varepsilon}$ , $\det(\mathbf{C}) = 1$ and neglecting $\nabla \mathbf{C}$ as well as higher order derivative terms of the right Cauchy Green tensor and all nonlinear terms, the linearized Euler-Lagrange equations may be written as,

$$\nabla \cdot \sigma + \mathbf{f} = \mathbf{0} \tag{75}$$

$$\partial_{\alpha} \left[ k_6 e_{\alpha\mu\nu} \varepsilon^{d}_{\mu\nu} + 9 k_6^2 K_{\alpha\beta} \mathrm{E}_{\beta} \right] = 0 \tag{76}$$

where,

$$\sigma_{\mu\nu} = \left( \mathbb{C}_{\mu\nu\alpha\beta} - \frac{1}{3} \mathbb{C}_{\sigma\sigma\alpha\beta} \mathrm{I}_{\mu\nu} \right) \varepsilon^{d}_{\alpha\beta} - 2k_6 \left( e_{\alpha\mu\nu} - \frac{1}{3} e_{\alpha kk} \mathrm{I}_{\mu\nu} \right) \mathrm{E}_{\alpha} \tag{77}$$

Comparing equations (71) and (72) with the linearized equations (75) and (76) via the space-time theory under isochoric deformation and equating the corresponding material constants coefficients, we get,

$$\mathbb{C} = \bar{\mathbb{C}}, \; e_{\alpha\mu\nu} = \frac{\bar{e}_{\alpha\mu\nu}}{2k_6} \; \text{ and } \; K_{\alpha\beta} = \frac{\bar{K}_{\alpha\beta}}{18 K_6^2} \tag{78}$$

Eliminating the time component from equations (56) and (58), we have,

$$\gamma^{+} = \frac{k}{c_{el}} \left( -\mathbf{E} + c_{el} \mathbf{B} \right) \tag{79}$$

$$\gamma^{-} = \frac{k}{c_{el}} \left( -\mathbf{E} - c_{el} \mathbf{B} \right) \tag{80}$$

From the dimensional analysis of equations (79) and (80) and using equations (63) and (64), we get, $k_6 = -\bar{\mu}\varepsilon_0$ and $k_7 = -k_6 c_{el}$, where $\bar{\mu}$ is a proportionality constant having units of meter per coulomb $\left(m/C\right)$. $c_{el}$ is given as, $c_{el} = c_0 \big/ \sqrt{\xi_r}$ where, $c_0$ is speed of light in free space and $\xi_r$ is the product of relative permittivity and relative permeability of the material.

### 6.3 Solution for isochoric deformation

We consider the following deformation map:

$$
\begin{aligned}
&\mathrm{x}_1 = \mathrm{X}_1 + \mathcal{P}\left(\mathrm{X}_2, t\right), \\
&\mathrm{x}_2 = \mathrm{X}_2, \\
&\mathrm{x}_3 = \mathrm{X}_3, \qquad \text{where, } \mathcal{P}\left(\mathrm{X}_2, t\right) = e^{i(k_2^f \mathrm{X}_2 + \omega^f t)},\ \omega^f = 50\,\mathrm{MHz}
\end{aligned}
\tag{81}
$$

$\mathrm{x}_i$ and $\mathrm{X}_j$, $i, j \in \left(1,2,3\right)$ are the coordinate covers of the current and the reference configurations respectively. For the given deformation map, the right Cauchy-Green tensor is,

$$
\mathbf{C} = \begin{pmatrix} 1 & \mathcal{P}' & 0 \\ \mathcal{P}' & 1+\mathcal{P}'^2 & 0 \\ 0 & 0 & 1 \end{pmatrix}; \quad \text{where, } \mathcal{P}' = \frac{\partial \mathcal{P}\left(\mathrm{X}_2, t\right)}{\partial \mathrm{X}_2}
\tag{82}
$$

Note that $\det\left(\mathbf{C}\right) = 1$ and so the given deformation is indeed isochoric. We choose a transversely isotropic piezoelectric material, PZT-5H. The material constants coefficients are given in Table 1 (Zhu *et al.*, 1998). Using Table 1 that furnishes the material constants, the material parameters for our theory are established following Section 6.2. We assume an infinite domain.

**Table 1**: Material parameters for PZT-5H

| $\bar{\mathbb{C}}_{1111}$ (GPa) | $\bar{\mathbb{C}}_{1122}$ (GPa) | $\bar{\mathbb{C}}_{1133}$ (GPa) | $\bar{\mathbb{C}}_{3333}$ (GPa) | $\bar{\mathbb{C}}_{2323}$ (GPa) |
|---|---|---|---|---|
| 113 | 67.3 | 66.9 | 98.9 | 20.1 |

| $\bar{e}_{113}\ \left(C/m^2\right)$ | $\bar{e}_{311}\ \left(C/m^2\right)$ | $\bar{e}_{333}\ \left(C/m^2\right)$ | $\bar{K}_{11}=\bar{K}_{22}\left(C^2/Nm^2\right)$ | $\bar{K}_{33}\left(C^2/Nm^2\right)$ |
|---|---|---|---|---|
| 14.8 | -5.26 | 23.5 | 1550 $\varepsilon_0$ | 1691 $\varepsilon_0$ |

Note that, in Table 1, the permittivity of free space, $\varepsilon_0 = 8.854\times10^{-12}\, C^2/Nm^2$ . The values of $\bar{\mu}$ and $\xi_r$ are assumed as $10^9\, m/C$ and 1500 respectively.

### 6.3.1 Equations of motion

We derive the Euler-Lagrange equations (65) and (66) using the given deformation (equation 81) and the PZT-5H material parameters. The Euler-Lagrange equation for the electric scalar potential may be written as,

$$18K_{\alpha\alpha}k_6^2\left[-\nabla_\alpha^2\varphi-\nabla_\alpha\left(\frac{\partial \mathrm{A}_\alpha}{\partial t}\right)\right]=0 \tag{83}$$

Similarly, the Euler-Lagrange equations for the magnetic vector potential along the three coordinate axes are presented below.

EoM along direction 1:

$$\begin{aligned}&-18K_{11}k_6^2\left[\partial_1\frac{\partial\varphi}{\partial t}+\frac{\partial^2\mathrm{A}_1}{\partial t^2}\right]+2k_7e_{322}\mathcal{P}'\mathcal{P}''\\&+18K_7^2\left[K_{22}\partial_3\partial_3\mathrm{A}_1+K_{33}\partial_2\partial_2\mathrm{A}_1-K_{22}\partial_3\partial_1\mathrm{A}_3-K_{33}\partial_2\partial_1\mathrm{A}_2\right]=0\end{aligned} \tag{84}$$

EoM along direction 2:

$$\begin{aligned}&-18K_{22}k_6^2\left[\partial_2\frac{\partial\varphi}{\partial t}+\frac{\partial^2\mathrm{A}_2}{\partial t^2}\right]+\\&18K_7^2\left[-K_{11}\partial_3\partial_2\mathrm{A}_3-K_{33}\partial_1\partial_2\mathrm{A}_1+K_{11}\partial_3\partial_3\mathrm{A}_2+K_{33}\partial_1\partial_1\mathrm{A}_2\right]=0\end{aligned} \tag{85}$$

EoM along direction 3:

$$-18K_{33}k_6^2\left[\partial_3\frac{\partial\varphi}{\partial t}+\frac{\partial^2\mathrm{A}_3}{\partial t^2}\right]+2k_6e_{322}\mathcal{P}'\frac{\partial^2\mathcal{P}}{\partial X_2\partial t}$$
$$+18K_7^2\left[K_{11}\partial_2\partial_2\mathrm{A}_3+K_{22}\partial_1\partial_1\mathrm{A}_3-K_{11}\partial_2\partial_3\mathrm{A}_2-K_{22}\partial_1\partial_3\mathrm{A}_1\right]=0 \tag{86}$$

### 6.3.2 Governing equations in Fourier space

We solve the Euler-Lagrange equations (83) through (86) in the Fourier space. $(\bar{\ })$ and $(\hat{\ })$ respectively denote Fourier transforms in space and time. The governing Euler-Lagrange equation for the electric scalar potential in Fourier domain is given as,

$$18k_6^2K_{\alpha\alpha}\left[\bar{k}_\alpha^2\,\hat{\bar{\varphi}}+\hat{\omega}\bar{k}_\alpha\,\hat{\bar{\mathrm{A}}}_\alpha\right]=0 \tag{87}$$

From equation (84), the Fourier transform of the equation for magnetic vector potential along direction 1 is,

$$-18k_6^2K_{11}\left[-\bar{k}_1\,\hat{\omega}\hat{\bar{\varphi}}-\hat{\omega}^2\,\hat{\bar{\mathrm{A}}}_1\right]+2e_{322}k_7\,\hat{\mathcal{F}}_t\left(\bar{\mathcal{F}}_{\boldsymbol{x}}\left((\mathcal{P})',(\mathcal{P})''\right)\right)$$
$$+18k_7^2\left[-\bar{k}_3^2K_{22}\hat{\bar{\mathrm{A}}}_1-\bar{k}_2^2K_{33}\hat{\bar{\mathrm{A}}}_1+\bar{k}_1\bar{k}_3K_{22}\hat{\bar{\mathrm{A}}}_3+\bar{k}_1\bar{k}_2K_{33}\hat{\bar{\mathrm{A}}}_2\right]=0 \tag{88}$$

Similarly, from equations (85) and (86), the Fourier transformed equations for magnetic vector potential along directions 2 and 3 are,

$$-k_6^2K_{22}\left[-\bar{k}_2\,\hat{\omega}\hat{\bar{\varphi}}-\hat{\omega}^2\,\hat{\bar{\mathrm{A}}}_2\right]$$
$$+k_7^2\left[-\bar{k}_3^2K_{11}\hat{\bar{\mathrm{A}}}_2-\bar{k}_1^2K_{33}\hat{\bar{\mathrm{A}}}_2+\bar{k}_2\bar{k}_3K_{11}\hat{\bar{\mathrm{A}}}_3+\bar{k}_1\bar{k}_2K_{33}\hat{\bar{\mathrm{A}}}_1\right]=0 \tag{89}$$

$$-18k_6^2K_{33}\left[-\bar{k}_3\,\hat{\omega}\hat{\bar{\varphi}}-\hat{\omega}^2\,\hat{\bar{\mathrm{A}}}_3\right]+2e_{322}k_6\,\hat{\mathcal{F}}_t\left(\bar{\mathcal{F}}_{\boldsymbol{x}}\left(\mathcal{P}',\partial_t\mathcal{P}'\right)\right)$$
$$+18k_7^2\left[-\bar{k}_2^2K_{11}\hat{\bar{\mathrm{A}}}_3-\bar{k}_1^2K_{22}\hat{\bar{\mathrm{A}}}_3+\bar{k}_2\,\bar{k}_3\,K_{11}\,\hat{\bar{\mathrm{A}}}_2+\bar{k}_1\bar{k}_3K_{22}\hat{\bar{\mathrm{A}}}_1\right]=0 \tag{90}$$

The Fourier transforms of the forcing terms in the governing equations are as follows,

$$
\begin{aligned}
f_{\mathrm{A}_1} &= \hat{\mathcal{F}}_t\left(\bar{\mathcal{F}}_{\boldsymbol{x}}\left((\mathcal{P})',(\mathcal{P})''\right)\right) \\
&= -2i\pi\left(k_2^f\right)^3 \delta\left(\bar{k}_2 - 2k_2^f\right)\delta\left(-2\omega^f + \hat{\omega}\right)
\end{aligned}
\tag{91}
$$

$$
\begin{aligned}
f_{\mathrm{A}_3} &= \hat{\mathcal{F}}_t\left(\bar{\mathcal{F}}_{\boldsymbol{x}}\left(\mathcal{P}',\partial_t\mathcal{P}'\right)\right) \\
&= -2i\pi\left(k_2^f\right)^2 \omega^f \delta\left(\bar{k}_2 - 2k_2^f\right)\delta\left(-2\omega^f + \hat{\omega}\right)
\end{aligned}
\tag{92}
$$

Here $f_{\mathrm{A}_1}$ and $f_{\mathrm{A}_3}$ are the forcing terms in the governing equations for magnetic vector potential along directions 1 and 3 respectively. $\bar{k}_\alpha$, $\alpha \in (1,2,3)$ and $\hat{\omega}$ are respectively the spatial and temporal frequencies. We solve equations (87) through (90) in the Fourier space for different spatial and temporal frequencies; $\bar{k}_1 = 30, 40, 60\ (\mathrm{MHz})$, $\bar{k}_2 = 2k_2^f$, $\bar{k}_3 = 40(\mathrm{MHz})$, $\hat{\omega} = 2\omega^f$. The electric scalar potential and magnetic vector potential spectra are shown in Figure 2. We observe multiple peaks of the scalar and vector potentials over a wide range of frequencies. For clarity of presentation, we remove data pertaining to relatively smaller peaks and thus imparting no significant information. Maximum peak value is observed for $\bar{k}_1 = 30\,\mathrm{MHz}$. It is evident that under certain loading frequencies, electric and magnetic potentials may increase manifold, with significant relevance to future studies on efficient energy harvesting.

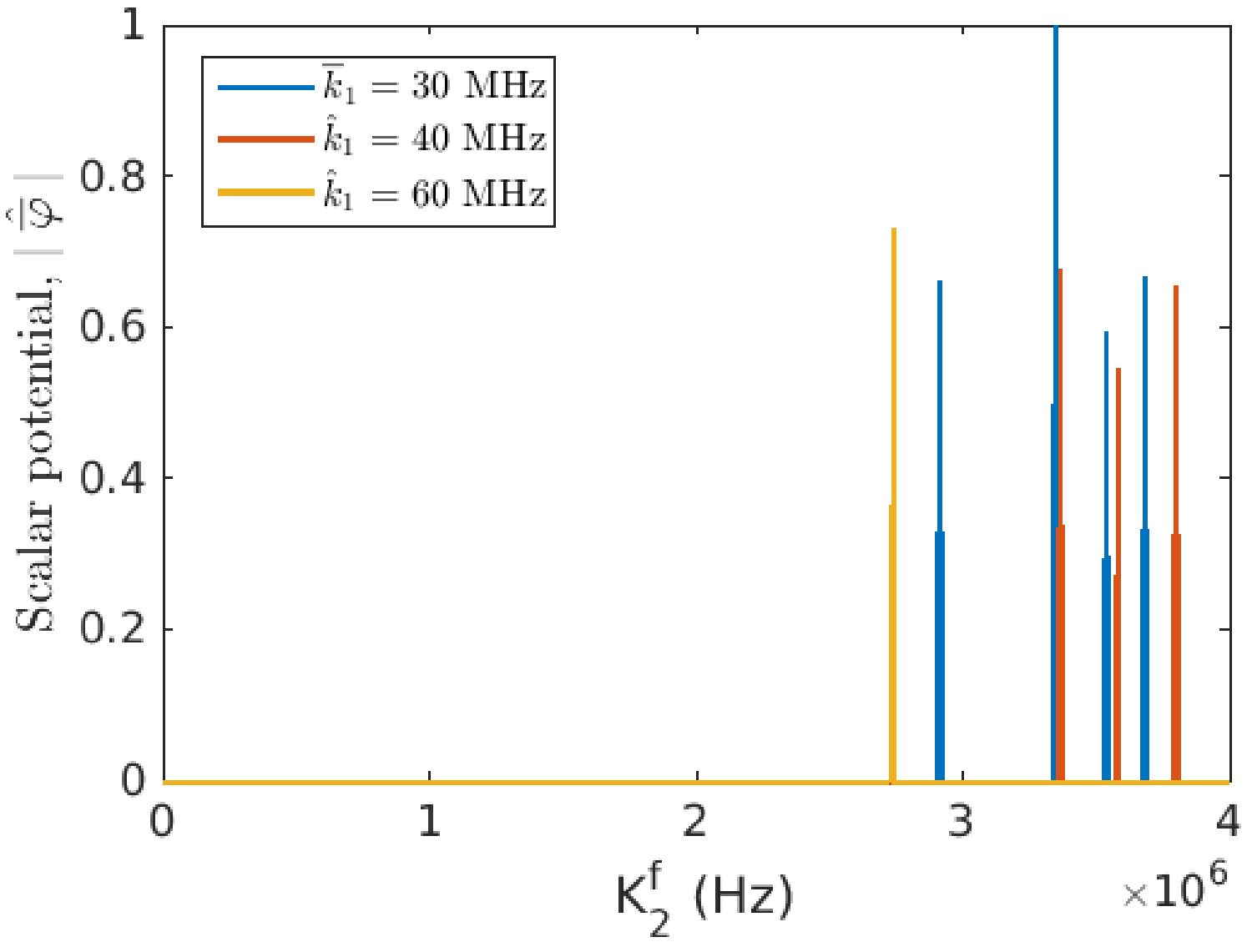

(a) Spectrum of electric scalar potential

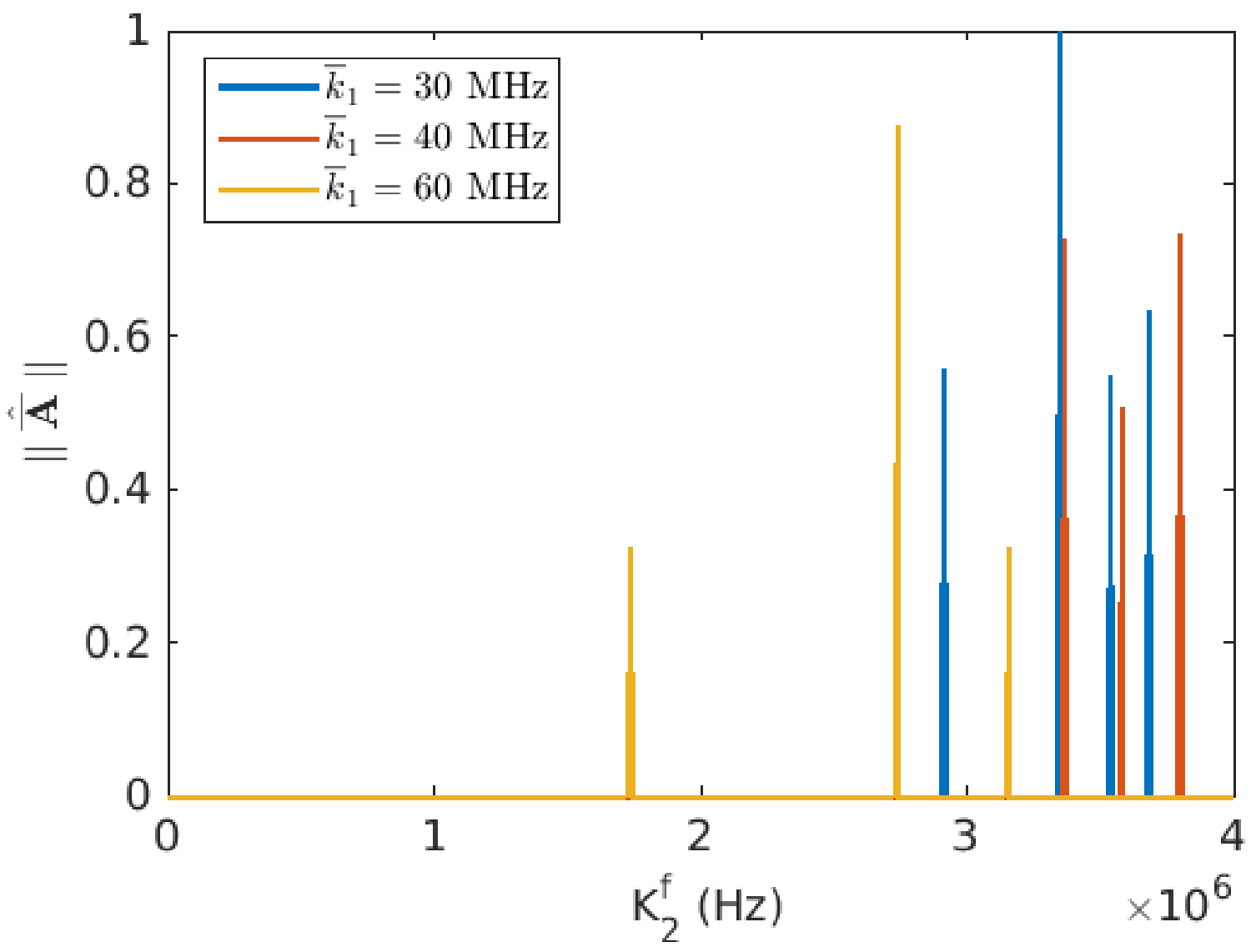

(b) Spectrum of the norm of magnetic vector potential

**Figure 2:** Electric scalar potential and magnetic vector potential spectra for different spatial Fourier frequency $\bar{k}_1$. Multiple peaks of the scalar and vector potentials over a wide range of frequencies are observed.

### 6.4 Piezoelectricity and piezomagnetism in a piezo-membrane

In this section, we solve the coupled Cauchy-Maxwell equations derived from our proposed theory for a piezo membrane subjected to tensile and transverse loading as shown in Figure 3 (see *Park* et al., 2017). The material of the membrane is polyvinylidene difluoride (PVDF). Dimensions of the membrane are d = 65 mm, b = 60 mm (see Figure 3) and its thickness is 80 μm. The top edge of the membrane is fixed and the bottom edge is kinematically constrained to the point of application of tensile loading.

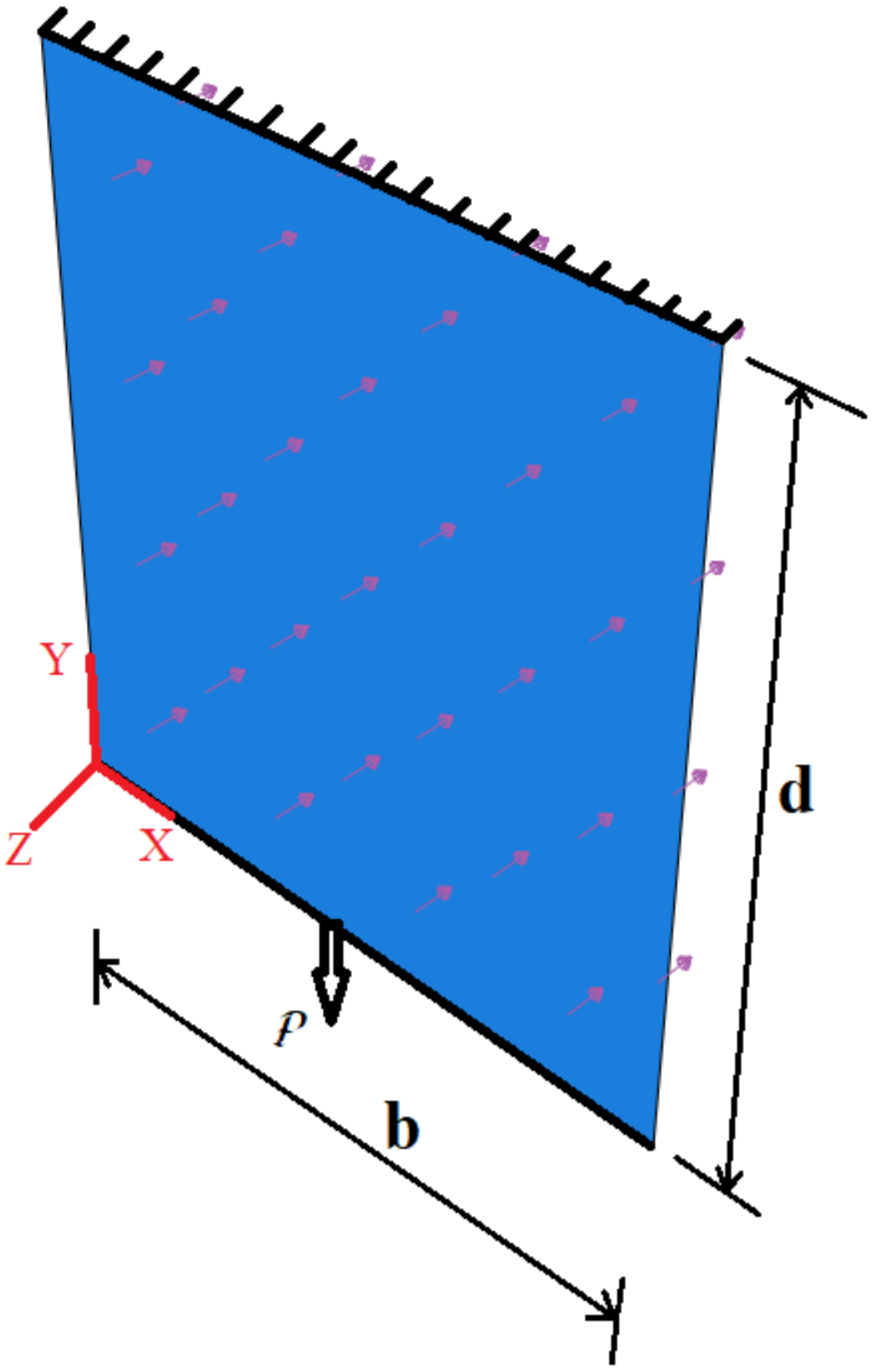

**Figure 3:** Schematic of the piezo-membrane

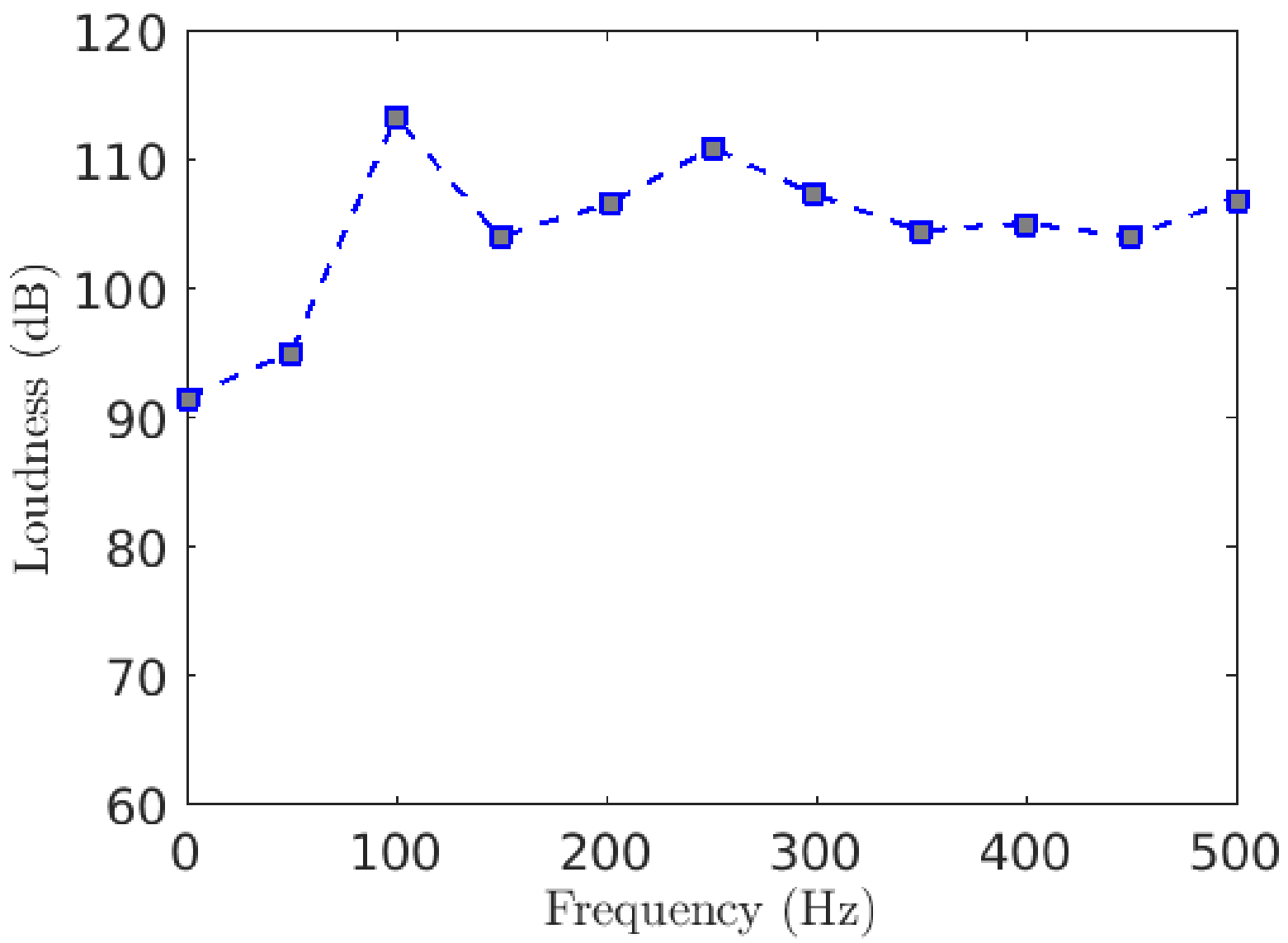

**Figure 2:** Transverse loading applied to the piezo-membrane (data for transverse loading is borrowed from Park *et al.*, 2017)

**Table 2:** Elastic stiffness coefficients for the PVDF membrane

| $\overline{\mathbb{C}}_{1111}$ (GPa) | $\overline{\mathbb{C}}_{1122}$ (GPa) | $\overline{\mathbb{C}}_{1133}$ (GPa) | $\overline{\mathbb{C}}_{3333}$ (GPa) | $\overline{\mathbb{C}}_{2222}$ (GPa) |
|---|---|---|---|---|
| 6.154 | 3.469 | 4.136 | 5.377 | 4.847 |
| $\overline{\mathbb{C}}_{2233}$ (GPa) | $\overline{\mathbb{C}}_{2323}$ (GPa) | $\overline{\mathbb{C}}_{1313}$ (GPa) | $\overline{\mathbb{C}}_{1212}$ (GPa) | |
| 3.508 | 0.104 | 0.104 | 0.544 | |

**Table 3:** Piezoelectric and dielectric constants for the PVDF membrane

| $\overline{e}_{113}\left(C/m^2\right)$ | $\overline{e}_{223}\left(C/m^2\right)$ | $\overline{e}_{311}\left(C/m^2\right)$ | $\overline{e}_{322}\left(C/m^2\right)$ | $\overline{e}_{323}\left(C/m^2\right)$ |
|---|---|---|---|---|
| -0.0028 | -0.0024 | 0.0414 | -0.0062 | -0.0584 |
| $\overline{K}_{11}\left(C^2/Nm^2\right)$ | $\overline{K}_{22}\left(C^2/Nm^2\right)$ | $\overline{K}_{33}\left(C^2/Nm^2\right)$ | $\varepsilon_0\left(C^2/Nm^2\right)$ | |
| $6.9\,\varepsilon_0$ | $8.6\,\varepsilon_0$ | $7.6\,\varepsilon_0$ | $8.854\times10^{-12}$ | |

We consider two cases of tensile load, viz. of 370 g and 1070 g; in both cases the membrane is subject to the same transverse loading as shown in Figure 3. The transverse loading profile is shown in Figure 4 (Park *et al.*, 2017). The material constants for the PVDF material are given in Tables 2 and 3 (Broadhurst *et al.*, 1984 and Wang *et al.*, 1993 ). For deriving the elastic stiffness matrix, the elastic compliance coefficients $\mathbb{S}_{44}$ and $\mathbb{S}_{66}$ are assumed close to the corresponding coefficients of P(VDF-TrFE). Specifically, the values of $\mathbb{S}_{44}$ and $\mathbb{S}_{66}$ are taken as $96.3\times10^{-10}\left(m^2/N\right)$ and $18.4\times10^{-10}\left(m^2/N\right)$ respectively. The mass density is 1780 $\mathrm{Kg}/m^3$, $\bar{\mu}=1e9\ \ m/C$ and $\xi_r=4.5e4$.

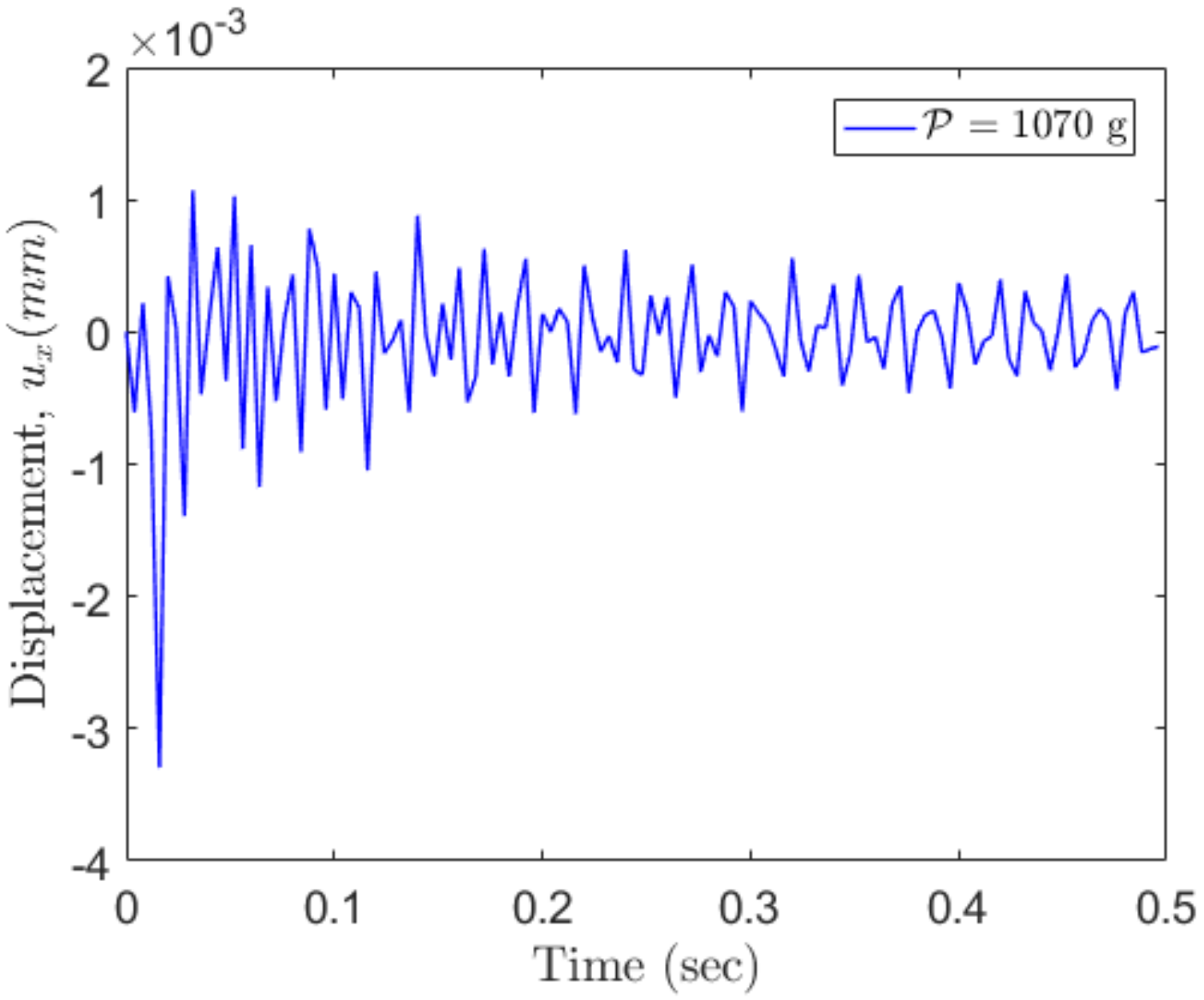

(a) X-displacement

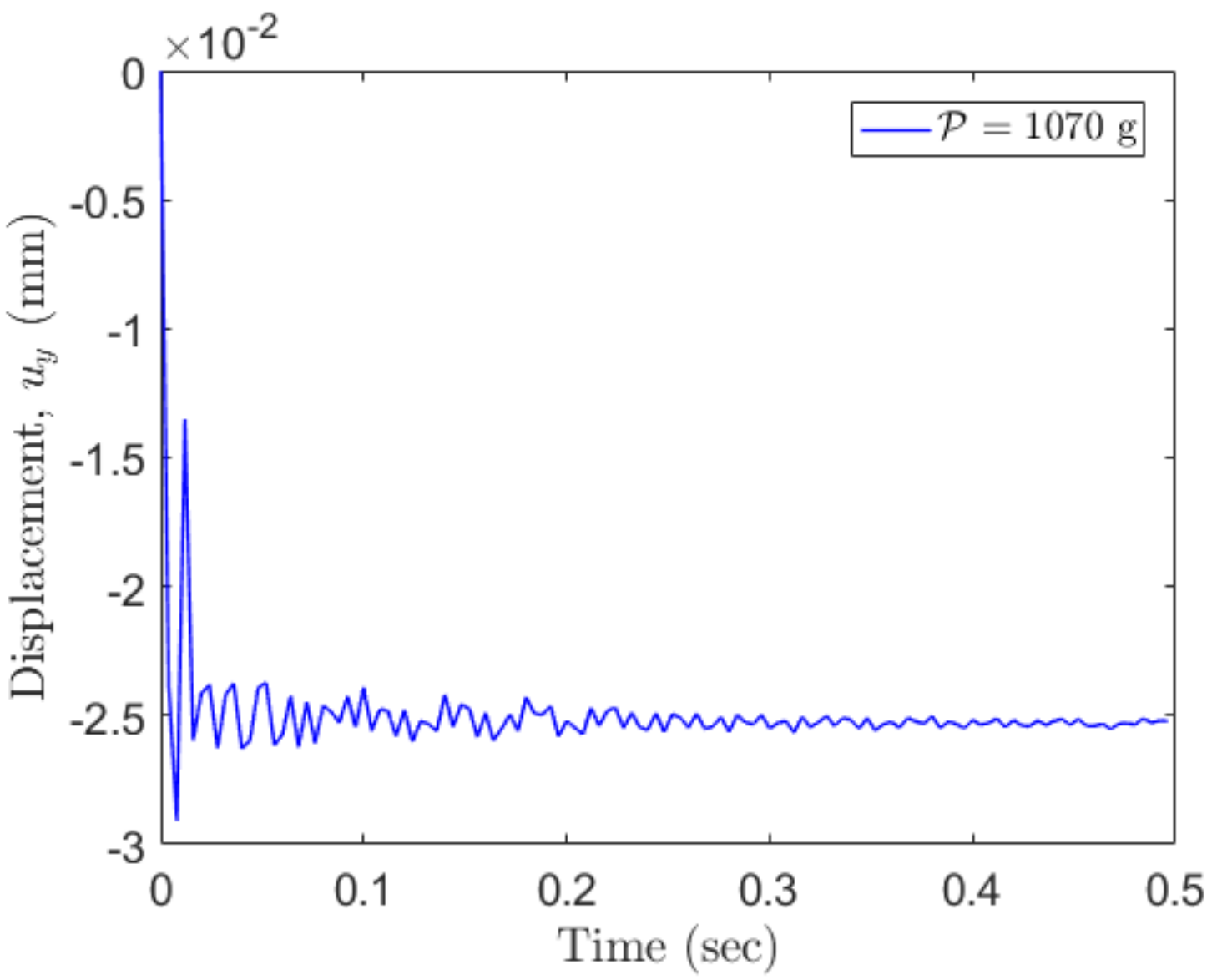

(b) Y-displacement

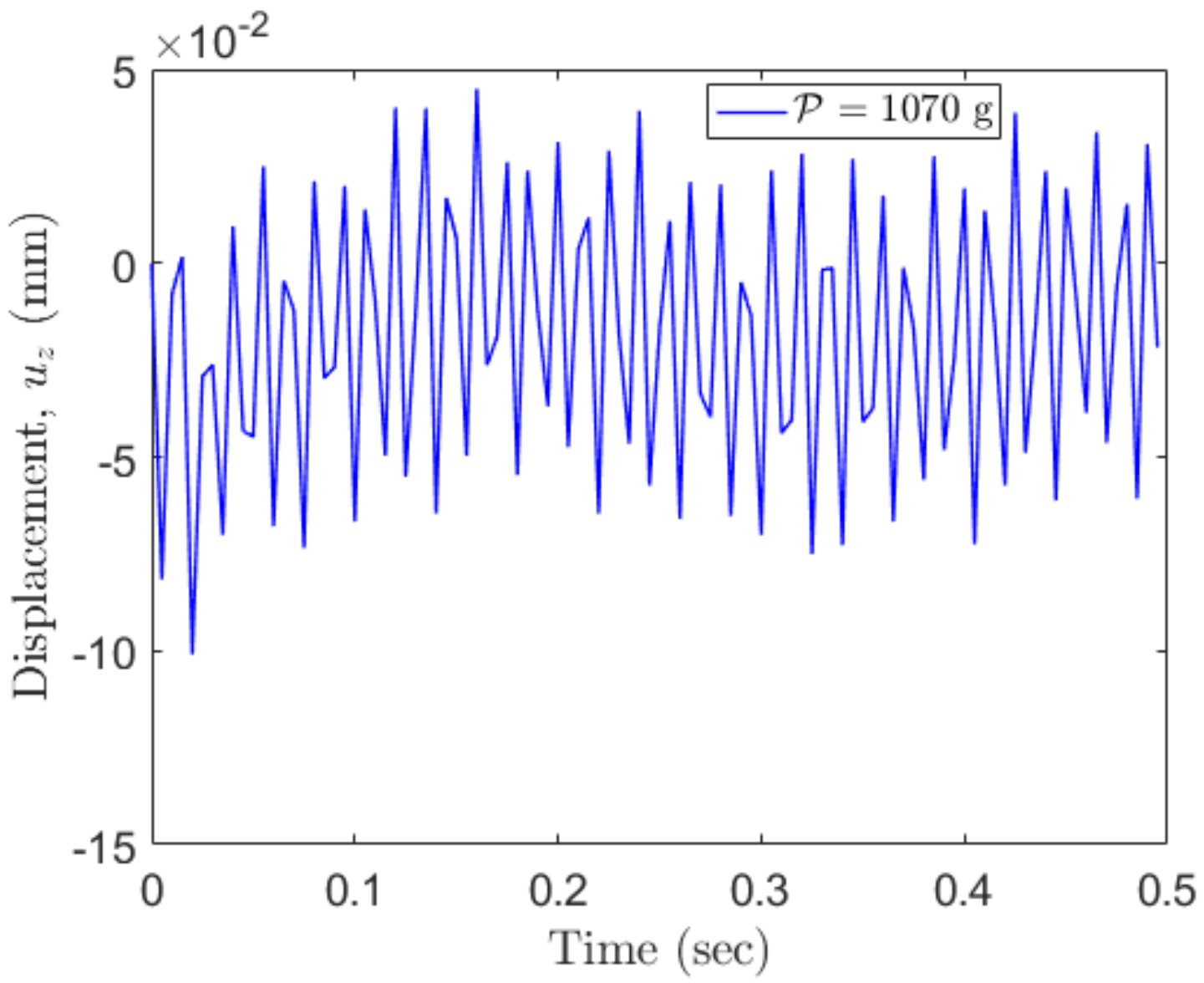

(c) Z-displacement

**Figure 5:** Displacemet time histories computed from uncoupled membrane elastic equations

We solve the governing Euler-Lagrange equations in a staggered way, i.e. we first solve for the deformation time history and with the known deformation history, we solve for the potentials in frequencies domain. Deformation time history is computed via the commercial finite element (FE) software ABAQUS 6.14®, where M3D4R elements are used for spatial discretization. Computation of the right Cauchy-Green tensor and its gradient from discrete deformation data using non-local peridynamics theory is explained briefly in Appendix A. Displacement time history at the central node is shown in Figure 5. The deformation field contours at different time instants, viz. $t = 0.1$, 0.3 and 0.5 (sec.), are shown in Figure 6. From the deformation contour plots, it is evident that corrugations along the Z-axis diminish as deformation reaches steady state with time.

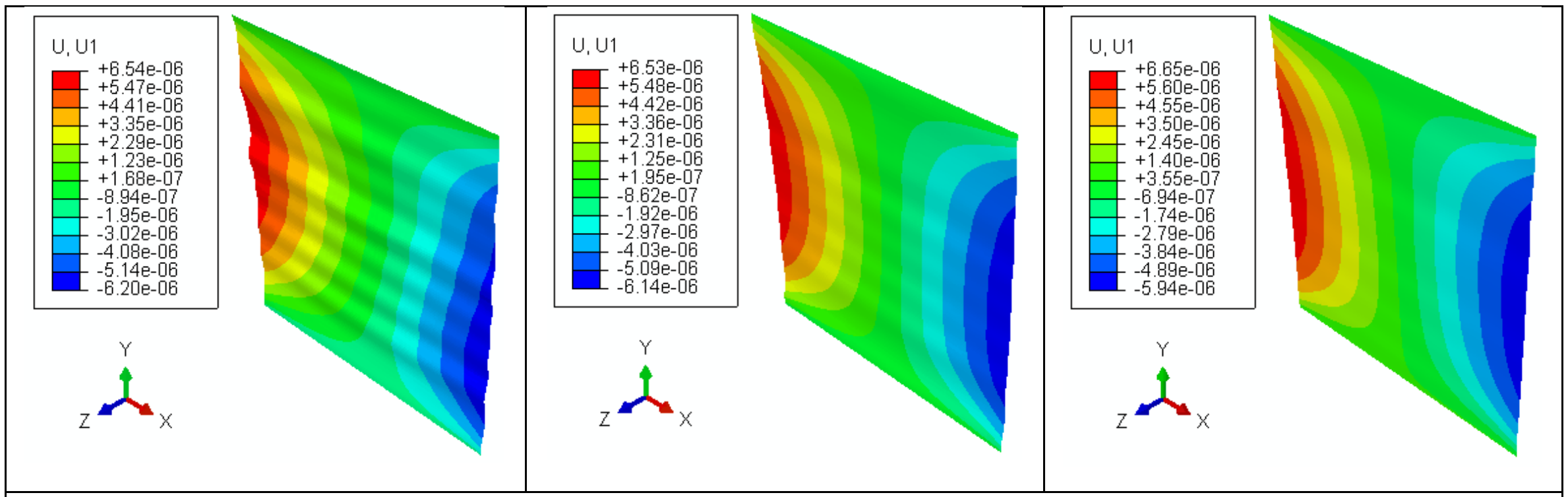

(a) X-component of displacement, $U_1$ $(m)$

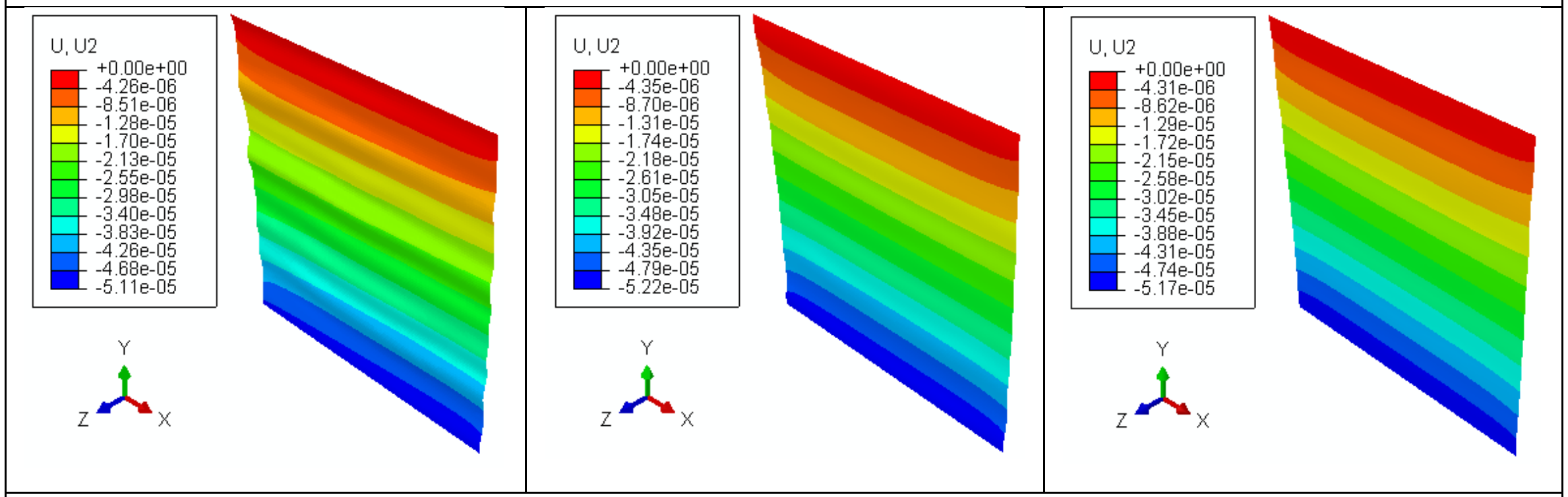

(b) Y-component of displacement, $U_2$ $(m)$

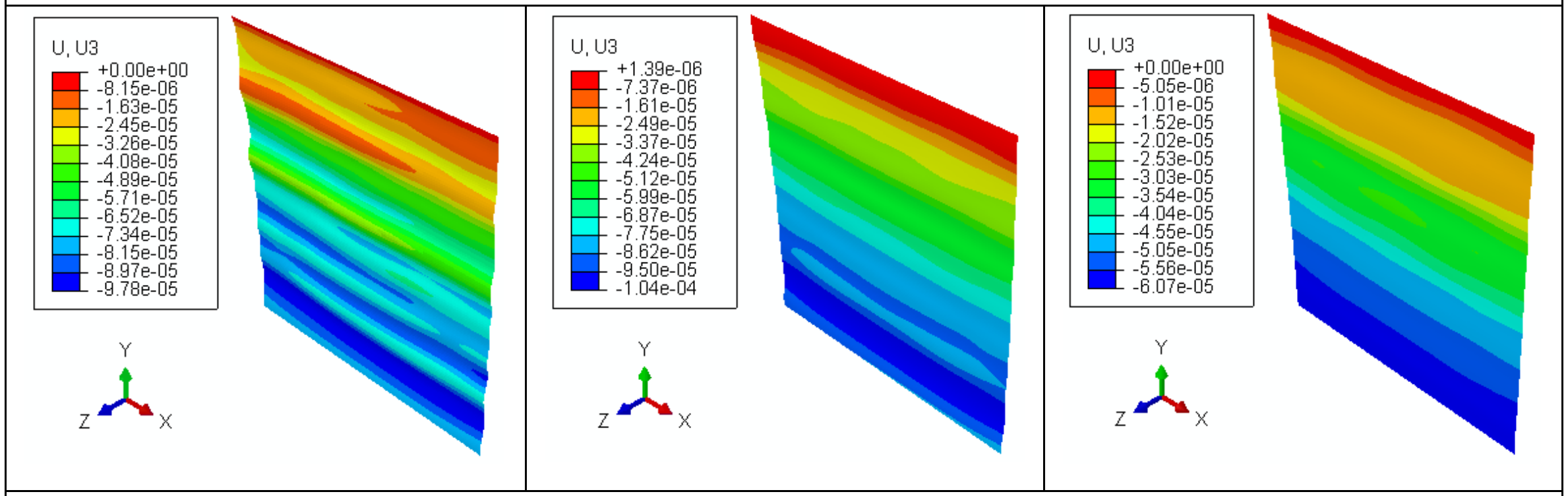

(c) Z-component of displacement, $U_3$ $(m)$

| Time (sec) | t = 0.1 | t = 0.3 | t = 0.5 |
|---|---|---|---|

**Figure 6:** Deformation contour plots at different time instants. Corrugations along the Z-axis diminish as the deformation reaches a steady state (magnification factor: 30)

The governing equations for electric scalar and magnetic vector potentials for the PVDF membrane can be derived by putting in the non-zero material constants in equations (65) and (66); see Section 6.1. Higher order temporal derivatives of deformation are neglected in the Euler-Lagrange equations. For electric scalar potential, the Euler-Lagrange equation is,

$$18K_{\alpha\alpha}k_6^2\left[-\nabla_\alpha^2\varphi-\nabla_\alpha\left(\frac{\partial \mathrm{A}_\alpha}{\partial t}\right)\right]=f_\varphi \tag{93}$$

where,

$$f_\varphi=\frac{2}{3}k_6\left(\det\mathbf{C}\right)^{-\frac{1}{3}}\left[e_{113}\,\partial_1\mathrm{C}_{13}+e_{223}\partial_2\mathrm{C}_{23}\right]-6K_{\alpha\alpha}\left(\partial_\alpha\mathrm{C}_{rs}^{-1}\right)\left(\partial_\alpha\mathrm{C}_{rs}\right)-6K_{\alpha\alpha}C_{rs}^{-1}\left(\partial_\alpha\partial_\alpha\mathrm{C}_{rs}\right) \tag{94}$$

Similarly, governing equations for the magnetic vector potential along the three axes are,

$$\begin{aligned}&-18K_{11}k_6^2\left[\partial_1\frac{\partial\varphi}{\partial t}+\frac{\partial^2\mathrm{A}_1}{\partial t^2}\right]\\&+18K_7^2\left[K_{22}\partial_3\partial_3\mathrm{A}_1+K_{33}\partial_2\partial_2\mathrm{A}_1-K_{22}\partial_3\partial_1\mathrm{A}_3-K_{33}\partial_2\partial_1\mathrm{A}_2\right]=f_{\mathrm{A}_1}\end{aligned} \tag{95}$$

$$\begin{aligned}&-18K_{22}k_6^2\left[\partial_2\frac{\partial\varphi}{\partial t}+\frac{\partial^2\mathrm{A}_2}{\partial t^2}\right]+\\&18K_7^2\left[-K_{11}\partial_3\partial_2\mathrm{A}_3-K_{33}\partial_1\partial_2\mathrm{A}_1+K_{11}\partial_3\partial_3\mathrm{A}_2+K_{33}\partial_1\partial_1\mathrm{A}_2\right]=f_{\mathrm{A}_2}\end{aligned} \tag{96}$$

$$\begin{aligned}&-18K_{33}k_6^2\left[\partial_3\frac{\partial\varphi}{\partial t}+\frac{\partial^2\mathrm{A}_3}{\partial t^2}\right]\\&+18K_7^2\left[K_{11}\partial_2\partial_2\mathrm{A}_3+K_{22}\partial_1\partial_1\mathrm{A}_3-K_{11}\partial_2\partial_3\mathrm{A}_2-K_{22}\partial_1\partial_3\mathrm{A}_1\right]=f_{\mathrm{A}_3}\end{aligned} \tag{97}$$

where, the forcing terms are given as,

$$f_{A_1} = -k_7\,\partial_2\left(\det \mathrm{C}\right)^{-\frac{1}{3}}\left[e_{322}\mathrm{C}_{22} + e_{311}\mathrm{C}_{11} + e_{333}\mathrm{C}_{33}\right] \tag{98}$$

$$f_{A_2} = k_7\,\partial_1\left(\det \mathrm{C}\right)^{-\frac{1}{3}}\left[e_{322}\mathrm{C}_{22} + e_{311}\mathrm{C}_{11} + e_{333}\mathrm{C}_{33}\right] \tag{99}$$

$$f_{A_3} = k_7 e_{113}\mathrm{C}_{13}\,\partial_2\left(\det \mathrm{C}\right)^{-\frac{1}{3}} - k_7 e_{223}\mathrm{C}_{23}\,\partial_1\left(\det \mathrm{C}\right)^{-\frac{1}{3}} \tag{100}$$

### 6.4.1 Solution in frequency domain

We solve the governing equations derived above in Fourier domain. The transformed equation for electric scalar potential is given as,

$$18k_6^2 K_{\alpha\alpha}\left[\bar{k}_\alpha^2\,\hat{\bar{\varphi}} + \hat{\omega}\,\bar{k}_\alpha\,\hat{\bar{\mathrm{A}}}_\alpha\right] = \hat{\bar{f}}_\varphi \tag{101}$$

Similarly, the transformed equations in the frequency domain for magnetic vector potential are,

$$\begin{aligned}&-18k_6^2 K_{11}\left[-\bar{k}_1\,\hat{\omega}\,\hat{\bar{\varphi}} - \hat{\omega}^2\,\hat{\bar{\mathrm{A}}}_1\right]\\ &+18k_7^2\left[-\bar{k}_3^2 K_{22}\hat{\bar{\mathrm{A}}}_1 - \bar{k}_2^2 K_{33}\hat{\bar{\mathrm{A}}}_1 + \bar{k}_1\bar{k}_3 K_{22}\hat{\bar{\mathrm{A}}}_3 + \bar{k}_1\bar{k}_2 K_{33}\hat{\bar{\mathrm{A}}}_2\right] = \hat{\bar{f}}_{A_1}\end{aligned} \tag{102}$$

$$\begin{aligned}&-k_6^2 K_{22}\left[-\bar{k}_2\,\hat{\omega}\,\hat{\bar{\varphi}} - \hat{\omega}^2\,\hat{\bar{\mathrm{A}}}_2\right]\\ &+k_7^2\left[-\bar{k}_3^2 K_{11}\hat{\bar{\mathrm{A}}}_2 - \bar{k}_1^2 K_{33}\hat{\bar{\mathrm{A}}}_2 + \bar{k}_2\bar{k}_3 K_{11}\hat{\bar{\mathrm{A}}}_3 + \bar{k}_1\bar{k}_2 K_{33}\hat{\bar{\mathrm{A}}}_1\right] = \hat{\bar{f}}_{A_2}\end{aligned} \tag{103}$$

$$\begin{aligned}&-18k_6^2 K_{33}\left[-\bar{k}_3\,\hat{\omega}\,\hat{\bar{\varphi}} - \hat{\omega}^2\,\hat{\bar{\mathrm{A}}}_3\right]\\ &+18k_7^2\left[-\bar{k}_2^2 K_{11}\hat{\bar{\mathrm{A}}}_3 - \bar{k}_1^2 K_{22}\hat{\bar{\mathrm{A}}}_3 + \bar{k}_2\,\bar{k}_3\,K_{11}\,\hat{\bar{\mathrm{A}}}_2 + \bar{k}_1\bar{k}_3 K_{22}\hat{\bar{\mathrm{A}}}_1\right] = \hat{\bar{f}}_{A_3}\end{aligned} \tag{104}$$

where, $\hat{\bar{f}}_\varphi$ and $\hat{\bar{f}}_{A_i}$, $i \in (1,2,3)$ are the transformed forcing terms. We solve equations (101) through (104) by setting the values of the special Fourier frequencies $\bar{k}_1$, $\bar{k}_2$ and $\bar{k}_3$ as 30, 20 and 45 (MHz) respectively. The frequency spectra of electric and magnetic vector potentials for two load cases (370 g and 1070 g) are shown in Figures 7 and 8 respectively. We clearly see one sharp peak for load case-II (1070 g) in the frequency range of 250-275 Hz, whereas no sharp peaks are observed for load case-I (370 g). This also shows a good qualitative agreement with the experimental results in Park *et al.* (2017). The ratio of the maximum peak values for the load case-II and load case-I from our simulation is around 3.83, which may be compared with the experimental value 2.52; see Park *et al.*, 2017. Indeed, most of the important features of the potential spectra, experimentally reported in the last cited article, emerge from our simulations too. The limited accuracy may perhaps be ascribed to our staggered solution strategy and also due to the unavailability of a few material constants for PVDF used in our simulation.

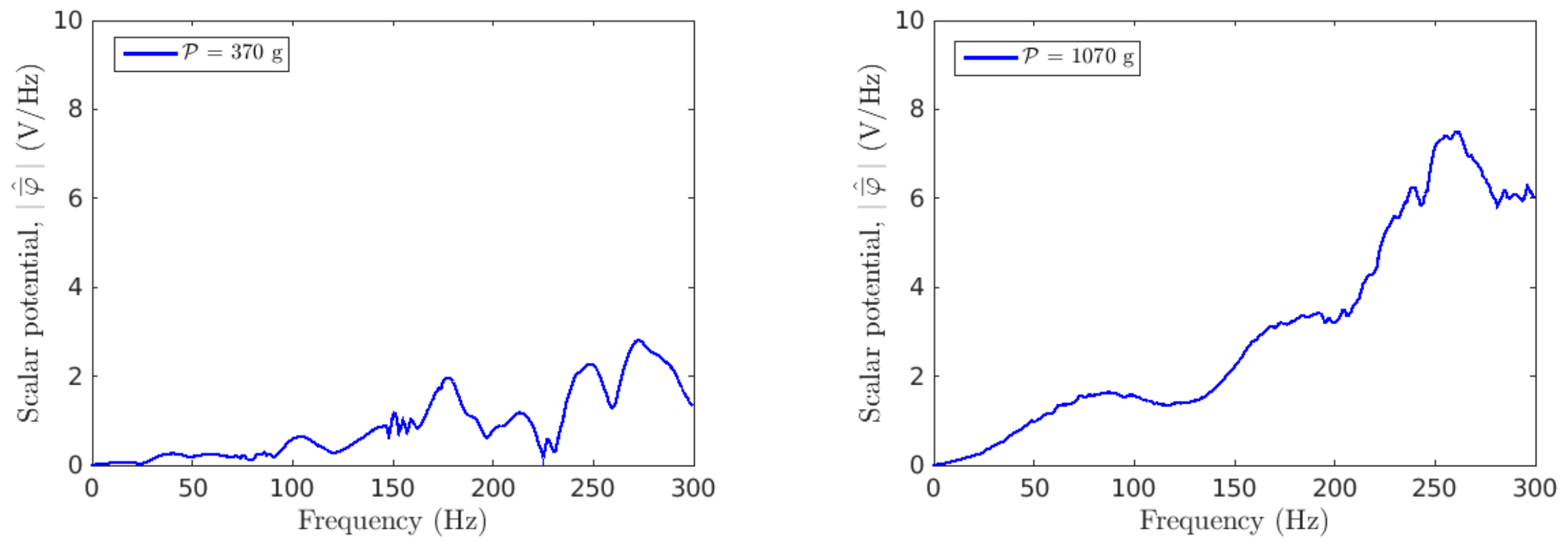

(a) (b)

(I) Electric scalar potential

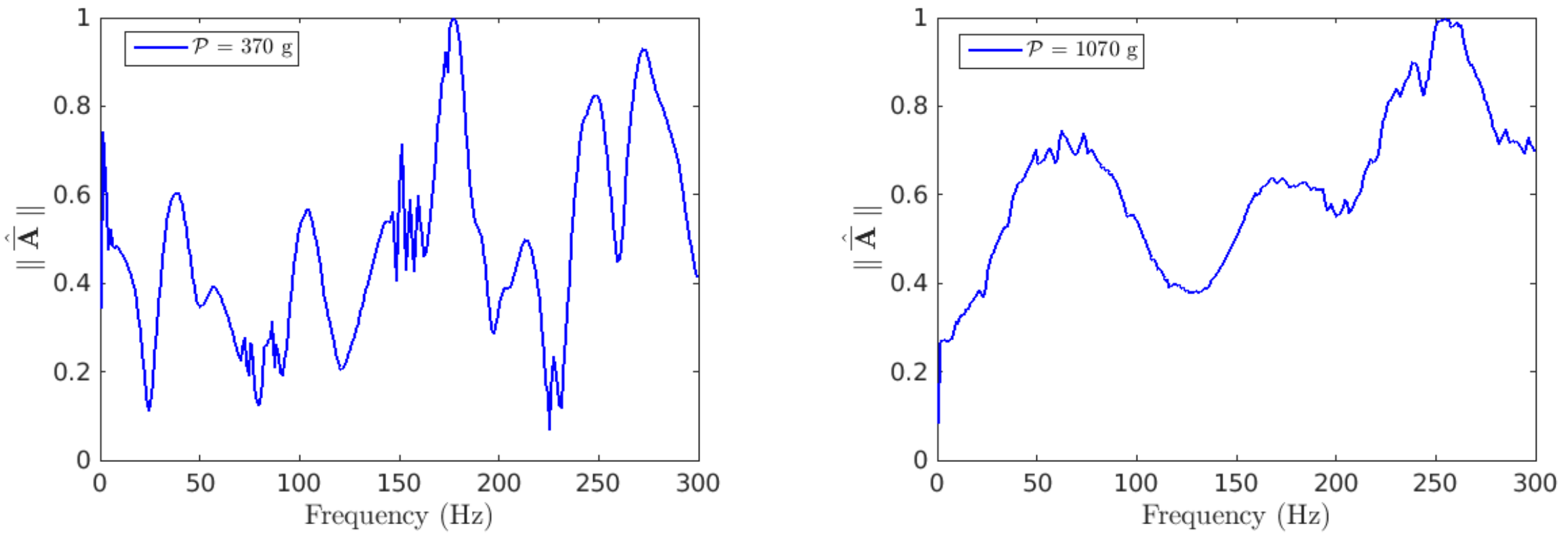

(c) (d)

(II) Magnetic vector potential (Norm)

**Figure 7:** Electric and magnetic vector potential spectra. (I) Electric scalar potential spectra, (a)

370 g, no sharp peak observed (b) 1070 g, sharp peak at (250-275) Hz (II) Magnetic vector potential (norm) spectra, (c) 370 g (d) 1070 g

## 7 Conclusion

Using a space-time conformal gauge theory of solids, we have tried to trace the origins of electro-magneto-mechanical coupling phenomena in solids purely based on certain symmetry arguments. Against a strictly finite deformation landscape, this has enabled us to combine or reconcile, for the first time to our knowledge, Cauchy's elasto-dynamic equations with Maxwell's equations of electromagnetism. To remain consistent with the classical ideas, we have recovered the energy density for Maxwell's equations for vacuum from our theory – an exercise that also marks a new derivation of Maxwell's equation. Thanks to the gauge connection, spatially and temporally higher order terms appear in the Lagrangian, leading to various pseudo-forces in the Euler-Lagrange equations. These terms may become significant under specific loading or geometric conditions and deserve a careful appraisal in a future work. With limited focus on only the electro-magneto-mechanical aspects in the response of solids, we have presently discussed the solution of a piezo-membrane under dynamic loading and derived specific conditions under which electric and magnetic potentials could be significantly amplified beyond what classical theories can predict. This observation could be of use in studies related to efficient energy harvesting. The space-time conformal gauge theory also affords a rational framework to understand wave propagation and dispersion in materials exhibiting electro-magneto-mechanical phenomena. Most importantly, extensions of the present work based on other forms of Lie group symmetries in space-time might pave the way for a better mathematical appreciation of defect-induced inelasticity in solids.

## Acknowledgement

This work is partially funded by the Indian Space Research Organization.

## Appendix A. Right Cauchy-Green tensor and its gradient from discrete deformation data

Consider a material body occupying a region $M_r \subset \mathbb{R}^3$ in the reference configuration and $M_c$ in the current configuration (see Figure 1). The coordinate cover for $M_r \subset \mathbb{R}^3$ is $\mathbf{X}$ and a deformation map $\boldsymbol{\chi}$ is defined through $\mathbf{y} = \boldsymbol{\chi}(\mathbf{X}, t)$, where $\mathbf{y}$ denotes the deformed position of material points in $M_c$. Also assume that we have numerically solved the governing equations so that function values of the deformation $\mathbf{y}^i = \boldsymbol{\chi}(\mathbf{X}^i, t)$ at discrete material points $\mathbf{X}^i$ ($i$ denotes the particle number) are available at any time. Our objective is to compute the right Cauchy-Green matrix ($\mathbf{C}$) and the gradients of various components of $\mathbf{C}$ from this discrete deformation data. In order to compute these, we employ a few concepts from peridynamics (PD). For details on peridynamics, we refer to Silling (2000) and Silling *et al*. (2007).

In non-ordinary state based PD, a method known as constitutive correspondence uses the classical constitutive equations through an equivalence between rates of strain energy densities from PD and classical continuum mechanics under homogeneous deformation conditions. There, we use a definition of nonlocal deformation gradient (an approximation of the deformation gradient) in the following manner:

$$\bar{\mathbf{F}}(\underline{\mathbf{Y}}) = \left[ \int_{\mathcal{H}} \omega(|\boldsymbol{\xi}|) \left( \underline{\mathbf{Y}}\langle \boldsymbol{\xi} \rangle \otimes \boldsymbol{\xi} \right) dV_{\mathbf{X}'} \right] \bar{\mathbf{K}}^{-1} \tag{A.1}$$

where, $\bar{\mathbf{K}} = \int_{\mathcal{H}} \omega(|\boldsymbol{\xi}|)(\boldsymbol{\xi} \otimes \boldsymbol{\xi}) dV_{\mathbf{X}'}$ is the shape tensor and $\omega(|\boldsymbol{\xi}|)$ the influence function. The domain of integration $\mathcal{H} \in M_r$ is known as the horizon which refers to a finite neighborhood over which a particle $\mathbf{X}$ interacts with its neighbors. The vector $\boldsymbol{\xi}=\mathbf{X}'\text{-}\mathbf{X}$ is referred to as a bond between particles $\mathbf{X} \in M_r$ and $\mathbf{X}' \in M_r$. $\underline{\mathbf{Y}}$ is known as a deformation vector state which is given by:

$$\underline{\mathbf{Y}}[\mathbf{X}]\langle \boldsymbol{\xi} \rangle = \boldsymbol{\chi}(\mathbf{X}') - \boldsymbol{\chi}(\mathbf{X}) \tag{A.2}$$

In the discretized setup, we may write the nonlocal deformation gradient $\bar{\mathbf{F}}$ at particle $\mathbf{X}^i$ as:

$$\bar{\mathbf{F}}\left(\mathbf{X}^i\right) \approx \left[\sum_{j\in\mathcal{H}^i} \omega\left(\left|\mathbf{X}^j - \mathbf{X}^i\right|\right)\left(\chi\left(\mathbf{X}^j\right) - \chi\left(\mathbf{X}^i\right)\right) \otimes \left(\mathbf{X}^j - \mathbf{X}^i\right)\Delta V^j\right]\bar{\mathbf{K}}^{-1}\left(\mathbf{X}^i\right) \tag{A.3}$$

where, $\bar{\mathbf{K}}\left(\mathbf{X}^i\right) \approx \sum_{j\in\mathcal{H}^i} \omega\left(\left|\mathbf{X}^j - \mathbf{X}^i\right|\right)\left(\mathbf{X}^j - \mathbf{X}^i\right) \otimes \left(\mathbf{X}^j - \mathbf{X}^i\right)\Delta V^j$. Having obtained $\bar{\mathbf{F}}\left(\mathbf{X}^i\right)$, we may compute the right Cauchy-Green tensor using $\mathbf{C}\left(\mathbf{X}^i\right) = \bar{\mathbf{F}}^T\left(\mathbf{X}^i\right)\bar{\mathbf{F}}\left(\mathbf{X}^i\right)$.

We also note that, for a scalar valued function $\varphi$, the nonlocal gradient ($\bar{\nabla}\varphi$) is defined through the following expression:

$$\bar{\nabla}\varphi\left(\mathbf{x}^i\right) \approx \bar{\mathbf{K}}^{-1}\left(\mathbf{x}^i\right)\left[\sum_{j\in\mathcal{H}_i} \omega\left(\left|\mathbf{x}^j - \mathbf{x}^i\right|\right)\left(\varphi\left(\mathbf{x}^j\right) - \varphi\left(\mathbf{x}^i\right)\right)\left(\mathbf{x}^j - \mathbf{x}^i\right)\Delta V^j\right] \tag{A.4}$$

Therefore, we may compute the gradients of various components of $\mathbf{C}\left(\mathbf{X}^i\right)$, for example, $\nabla \mathrm{C}_{11}$ by using $\mathrm{C}_{11}$ in place of $\varphi$ in equation (A.4).